\documentclass{pasj00}
\draft

\begin{document}
\SetRunningHead{L. -X. Li}{Energetics of Black Hole-Accretion Disk System ...}
\Received{2004/03/04}
\Accepted{2004/06/14}

\title{Energetics of Black Hole-Accretion Disk System with Magnetic Connection: 
Limit of Low Accretion Rate}

\author{Li-Xin \textsc{Li}%
  \thanks{Chandra Fellow}}
\affil{Harvard-Smithsonian Center for Astrophysics, Cambridge, MA 02138, USA}
\email{lli@cfa.harvard.edu}


\maketitle

\begin{abstract}
We study the energetics of a black hole-accretion disk system with magnetic
connection: a Keplerian disk is connected to a Kerr black hole by a large-scale 
magnetic field going through the transition region. We assume that the magnetic 
field is locked to the inner boundary of the disk and corotates with the inner 
boundary, the accretion rate is low but the accretion from the disk can still 
provide enough amount of cold plasma particles in the transition region so that 
the magnetohydrodynamics approximation is valid. Then, the magnetic field is 
dynamically important in the transition region and affects the transportation
of energy and angular momentum. Close to the equatorial plane, the motion of
particles is governed by a one-dimensional radial momentum equation, which 
contains a fast critical point as the only intrinsic singularity. By finding 
solutions that smoothly pass the fast critical point, we find that a system 
with a fast rotating black hole and that with a slow rotating black hole behave 
very differently. For a black hole with $a > a_{\rm cr}\equiv 0.3594 M$, where 
$M$ is the mass, $a$ the specific angular momentum of the black hole, the 
spinning energy of the black hole is efficiently extracted by the magnetic field 
and transported to the disk, increasing the radiation efficiency of the disk by 
many orders of magnitude. For a black hole with $0\leq a < a_{\rm cr}$, the 
inner region of the disk is disrupted by the magnetic field and the inner 
boundary of the disk moves out to a radius where the Keplerian angular velocity 
of the disk is equal to the spinning angular velocity of the black hole (which 
is at infinity if the 
black hole is nonrotating). As a result, the disk may have an extremely low
radiation efficiency if $0\leq a/M \ll 1$. Although our calculations are 
restricted in a thin slab close to the equatorial plane, we expect that the 
solutions are typical for the whole transition region. 
\end{abstract}

\section{Introduction}
\label{sec1}

Extraction of energy from a rotating black hole with the aid of magnetic fields
has been investigated by many people for several decades. When a magnetic field
connects the horizon of a Kerr black hole to remote plasma particles, Blandford 
and Znajek (\yearcite{bla77}) showed that the voltage drop on the horizon of the 
black hole induced by the rotation of the black hole in the magnetic field is 
large enough to cause a cascade production of electron-positron pairs to form a 
force-free plasma around the black hole. As a result, a poloidal electric 
current propagating between the horizon of the black hole (the ``battery'') and 
the remote plasma (the ``load'') is produced, which transports energy and 
angular momentum from the black hole to the remote plasma in the form of 
Poynting flux. 
The Blandford-Znajek mechanism was later reformulated and confirmed by 
Macdonald and Thorne (1982) in terms of the ``membrane paradigm'' (see also 
\cite{tho86}). Although the original Blandford-Znajek mechanism was 
formulated with the assumption of a force-free plasma around the black hole, 
\citet{tak90} showed that the Blandford-Znajek mechanism also works for 
an ideal magnetohydrodynamical (MHD) fluid that is not force free. Under some
conditions the plasma accreting onto the black hole carries a negative 
energy flux, and as a result the rotational energy of the black hole is 
extracted. For a long time the Blandford-Znajek mechanism has been considered
as a promising process for powering the radio jets in active galactic nuclei
\citep{bla77,ree82,phi83,beg84}. Recently, this mechanism has also been 
invoked in models for the central engines of gamma-ray bursts 
\citep{pac93,mes97,pac98,lee00,li00b}.

Recently, a variant of the Blandford-Znajek mechanism has been discussed in
the literature: The magnetic field is assumed to connect the black hole to the 
accretion disk rotating around it \citep[and references 
therein]{bla99,li00a,li00c,put01,li02a,bla02,mik02,wan02,wan03,uzd03}. In 
such a model, the accretion disk replaces the remote plasma (the ``load'') in 
the standard Blandford-Znajek mechanism. Moreover, because of the rotation of
the disk, the disk itself is also a ``battery'', in addition to the black hole
battery. The two batteries have opposite signs and the direction of the 
transportation of energy and angular momentum is determined by the net effect 
of the two batteries \citep{li02a}. When the black hole rotates faster than the 
disk, its spin energy 
is extracted by the magnetic field and transported to the disk region. This 
energy is eventually dissipated and radiated away by the disk, increasing the 
power and the radiation efficiency of the disk \citep{ago00,li00a,li02a}. When 
the black hole rotates slower than the disk, the magnetic field transports 
energy and angular momentum from the disk to the black hole, decreasing the
power and the radiation efficiency of the disk. Therefore, magnetic 
fields connecting a black hole to an accretion disk have important effects on 
the transportation of energy and angular momentum, and the radiation process 
in the disk. Recent {\em XMM-Newton} observations on some Seyfert galaxies and 
Galactic black hole candidates have revealed some possible evidences for a 
magnetic connection between a black hole and its disk and extraction of energy 
from a Kerr black hole \citep{wil01,li02a,li02b,mil02,bal03,mer03,rey03,rey04}.

For a standard geometrically thin Keplerian disk, it is usually assumed that 
the inner boundary of the disk is located at the marginally stable circular 
orbit and the torque at the inner 
boundary is zero \citep{sha73,nov73,pag74}. This assumption has been justified
only for nonmagnetized or weakly magnetized disk flows \citep{muc82,abr89}. 
When the disk is strongly magnetized, the magnetic field may strongly affect 
the structure and accretion of the disk flow 
\citep{cam90,nit91,hir92,pun01,li03b,li03c}. In particular, the assumption of 
zero torque at the inner boundary of the disk has recently been challenged by 
\citet{kro99}, \citet{gam99}, and \citet{ago00}. These authors argued that, when 
magnetic fields are present, they may be dynamically important in the plunging 
region between the marginally stable circular orbit and the horizon of the black 
hole and couple the material in the plunging region to the inner boundary of 
the disk. As a result, the material in the plunging region exerts a torque to
the disk at the inner boundary. The issue is still under debate, however 
\citep{pac00,arm01,haw02,afs02,li03a,li03b}. For instance, \citet{li03b} 
argued that small-scale and tangled magnetic fields---which are often involved 
in MHD disks---can never be dynamically important. In the presence of an even 
small but nonzero resistivity, magnetic reconnection will work efficiently to 
limit the growth of the magnetic field (see also \cite{bla00}), so that in the 
equilibrium state the energy density of the magnetic field is always much 
smaller than the kinetic energy density of disk particles. This means that 
small-scale and chaotic magnetic fields in the plunging region cannot produce 
a significant torque at the inner boundary of the disk. 

However, large-scale and ordered magnetic fields have a completely different 
story. They are more easily amplified by the shear rotation of the disk. When 
the resistivity is small, in the equilibrium state the energy density of the 
magnetic field can be large enough to be equipartitioned with the kinetic 
energy density of disk particles, then the magnetic field must be dynamically 
important and affect the structure and the motion of the accretion flow 
\citep{li03b,li03c}. 

If in the model of \citet{gam99} the magnetic field is interpreted as a 
large-scale magnetic field connecting a Kerr black hole to a disk around it 
and corotating with the disk inner boundary, then Gammie's results showed 
that the magnetic field is dynamically important in the transition region, 
extracts the spin energy of the black hole and transports it to the disk. 
However, he has obtained a disk radiation efficiency that is only slightly 
bigger than unity: $\varepsilon_{\rm total} \approx 1.04$ for $a/M = 0.95$, 
where $\varepsilon_{\rm total}$ is defined by the ratio of the total 
radiated energy to the mass accretion rate, $M$ is the mass of the black 
hole, $a$ is the specific spinning angular momentum of the black 
hole.\footnote{Throughout the paper we use geometrized units $G = c =1$.}  
This is caused by the fact that Gammie has assumed that the strength of 
magnetic field in the transition region is limited by the disk accretion rate: 
$F_{\theta\phi} \approx r^2 B_r \approx 3 \eta r_{\rm g}^{1/4} r_{\rm in}^{3/4}$ 
[eq.~(12) of \cite{gam99}, where $r_{\rm g} = M$ is the gravitational radius 
of the black hole, $r_{\rm in}$ is the radius of the disk inner boundary, 
$\eta$ is a factor $\lesssim 1$]. However, this relation
crucially depends on the validity of the following steady state equation for a 
standard Keplerian accretion disk (with zero torque at the inner boundary)
\begin{eqnarray} 
     3\pi \Sigma \nu \approx \dot{M}_{\rm D},
	\label{vis1}
\end{eqnarray}
where $\Sigma$ is the surface mass density of the disk, $\nu$ is the disk 
viscosity, and $\dot{M}_{\rm D}> 0$ is the mass accretion rate of the disk. 
As we will see, equation~(\ref{vis1}) is not valid any more when there exists a 
large-scale magnetic field connecting a black hole to a disk.

A large-scale magnetic field connecting a Kerr black hole to a disk at the
inner boundary will produce a nonzero torque at the inner boundary. Then, 
in the steady state the internal viscous torque in the disk is given by 
\citep{li02a}
\begin{eqnarray}
     g = \frac{\left(E -\Omega_{\rm D} L\right)_{\rm in}}{E - \Omega_{\rm 
          D}L}\,g_{\rm in} + \frac{E - \Omega_{\rm D}L}{-d\Omega_{\rm D}/
          dr}\, \dot{M}_{\rm D} f,
	\label{vtorque}
\end{eqnarray}
where $E$ and $L$ are the specific energy and specific angular momentum of
disk particles on Keplerian circular orbits, $\Omega_{\rm D}$ is the disk
angular velocity, the function $f$ is given by equation~(15n) of \cite{pag74}, 
and the subscript ``in'' means evaluation at the inner boundary of the disk 
(i.e., the marginally stable circular orbit). Apparently, when $g_{\rm in}
\neq 0$, $\dot{M}_{\rm D}$ does not have to be nonzero to balance the 
internal viscous torque $g$. In the limiting case $\dot{M}_{\rm D} = 0$, 
the viscous torque in the disk just works to transport outward the angular 
momentum that is pumped into the disk by the magnetic connection.

Using $g\approx 3\pi r^2 \Sigma\Omega_{\rm D}\nu$, $E -\Omega_{\rm D} L 
\approx 1$, $-d\Omega_{\rm D}/dr \approx 3\Omega_{\rm D}/(2r)$, and $f
\approx (3/2) r \Omega_{\rm D}^2$, equation~(\ref{vtorque}) then becomes
\begin{eqnarray}
     3\pi \Sigma\nu \approx \frac{g_{\rm in}}{r^2 \Omega_{\rm D} }
	     + \dot{M}_{\rm D}.
	\label{vis2}
\end{eqnarray}
Equation~(\ref{vis2}) is an extension of the standard equation~(\ref{vis1}) to 
the case when the torque at the inner boundary of the disk, $g_{\rm in}$, is 
nonzero. Because of equation~(\ref{vis2}) [or, the equivalent 
equation~(\ref{vtorque})], the 
limit on $F_{\theta\phi}$ derived by Gammie does not apply to the case of a
large-scale magnetic field connecting a black hole to a disk. The relaxation
of this restriction allows us to consider the limit of low accretion rate 
and look for solutions with higher total radiation efficiency: $\varepsilon_{\rm 
total} \gg 1$.

In this paper, we extend the works of \citet{gam99} and Li (\yearcite{li03a,li03b}). 
We assume that a large-scale magnetic field connects a 
Kerr black hole to an accretion disk through the transition region between
the marginally stable circular orbit and the horizon of the black hole. In
addition, we assume that the mass accretion rate is low, but the accretion 
from the disk can still provide sufficient amount of plasma particles in the 
transition region so that the MHD approximation is valid.\footnote{This 
assumption means that the plasma in the transition region is nearly in a 
force-free condition (i.e., the electromagnetic forces dominate over the 
inertial and gravitational forces), except very close to horizon of the black 
hole.} The 
magnetic field is frozen to the perfectly conducting disk at the inner 
boundary and corotates with the inner boundary. We solve the one-dimensional 
radial momentum equation, derived by \citet{li03b} for the region close to
the equatorial plane (but we expect the solutions also apply, at least 
qualitatively, to the regions well above and below the equatorial plane), for 
smooth solutions 
corresponding to accretion flows starting from the inner boundary of the disk 
and ending on the horizon of the black hole. Based on these solutions, we 
study the effects of the magnetic field on the energetics of black 
hole-accretion disk system with magnetic connection, including extraction of 
energy from a rapidly rotating black hole and the influence on disk radiation 
efficiency.

\section{Outline of the Model and the Mathematical Formalism}
\label{sec2}

The geometry of the model is as follow: A large-scale 
magnetic field connects a Kerr black hole to a directly rotating Keplerian 
accretion disk, going through the transition region between the inner 
boundary of the disk and the horizon of the black hole. The mass accretion
rate is assumed to be low, resulting a sharp density contrast between the
disk region and the transition region. The high-density disk region is 
geometrically thin, presumably because the magnetic field in the disk region 
is not dynamically important (i.e., the energy density of the magnetic field 
is much smaller than the rotational energy density of disk particles). In the 
low-density transition region, the magnetic field must be dynamically important 
due to the low mass density of particles, so both magnetic fields and gases are 
geometrically thick and they expand significantly in the vertical direction. 
Possible formation scenarios for this model will be discussed in Sec.~\ref{sec4}.

We will focus on a small neighborhood of the equatorial plane in the transition 
region: $\cos^2\theta \ll 1$.\footnote{Throughout the paper we adopt the 
Boyer-Lindquist coordinates ($t,r,\theta,\phi$). The condition $\cos^2\theta \ll 1$
will not restrict the applicability of our results too much, see the discussions
in Sec.~\ref{sec4}.} Following \citet{gam99} and 
Li (\yearcite{li03a,li03b}), we assume that the plasma in the transition region 
is cold and perfectly conducting, near the equatorial plane the magnetic field 
and the velocity field of plasma particles have only $r$- and $\phi$-components. 
The mathematics for such a model is described in detail by Li (\yearcite{li03a,li03b}) 
(see also \cite{gam99,cam86a,cam86b,tak90,hir92,pun01,tak02}), and some results
are tested by numerical simulations \citep{gam03,vil03}. In summary, in the 
stationary and axisymmetric state, the motion of the plasma particles is 
governed by the one-dimensional radial momentum equation $F(r, u^r) = 0$, 
where the generation function $F(r, u^r)$ is defined by \citep{li03b}
\begin{eqnarray}
     F\left(r,u^r\right) &\equiv&
          \left\{1+ \frac{c_0^2}{r^2 u^r}\left[\chi^2 - \frac{A}{r^2}\left(
		\omega-\Omega_\Psi\right)^2\right]\right\}^2
          \left[\frac{r^2}{\Delta}\left(u^r\right)^2 + 1- 
          \frac{f_{\rm E}^{\prime 2}}{\chi^2 
		- \frac{A}{r^2}\left(\omega-\Omega_\Psi\right)^2}\right]
          \nonumber\\
          &&+\; \frac{\left\{\frac{A}{r^2}\left(\omega-\Omega_\Psi\right)
		f_{\rm E}^\prime+\left[\chi^2 - \frac{A}{r^2}\left(\omega-
		\Omega_\Psi\right)^2\right]f_{\rm L}\right\}^2}{\Delta\left[\chi^2 
		- \frac{A}{r^2}\left(\omega-\Omega_\Psi\right)^2\right]},
        \label{funcf}
\end{eqnarray}
where $\Delta \equiv r^2 -2 M r +a^2$, $A\equiv (r^2+a^2)^2 -\Delta a^2$, 
$\chi\equiv (r^2 \Delta/A)^{1/2}$ (the lapse function), $\omega\equiv 2Mar/A$
(the angular velocity of frame dragging), and $u^r$ is the radial component of 
the four-velocity of plasma particles. 

The definition of $F(r, u^r)$ contains four integral constants: $c_0\equiv
C_0/\sqrt{-F_{\rm m}}$, $C_0$ is the radial magnetic flux, $F_{\rm m} <0$ is 
the radial mass flux; $\Omega_\Psi$ corresponds to the ``angular velocity of
the magnetic field'' (see below for a precise definition for the angular 
velocity of magnetic field lines); $f_{\rm L} \equiv -F_{\rm L}/
F_{\rm m}$, $F_{\rm L}$ is the radial angular momentum flux; $f_{\rm E}^\prime
\equiv f_{\rm E} - \Omega_\Psi f_{\rm L}$, where $f_{\rm E} \equiv -F_{\rm E}/
F_{\rm m}$, $F_{\rm E}$ is the radial energy flux (for details see 
\cite{li03b}). Note, our $c_0$ corresponds to Gammie's $F_{\theta\phi}/
\sqrt{2}$, $\Omega_\Psi$ corresponds to his $\Omega_F$. (Since $F$ depends on
$c_0^2$, without loss of generality we assume $c_0>0$.)

We remark that the angular velocity of magnetic field lines---which is often 
defined for a degenerate electromagnetic field where the electric field is 
perpendicular to the magnetic field---and the angular velocity of disk particles 
are two different concepts. The angular velocity of magnetic field lines, 
$\Omega_{\rm F}$, is defined by 
\begin{eqnarray}
     F_{ab} \left[\left(\frac{\partial}{\partial t}\right)^b +
	     \Omega_{\rm F} \left(\frac{\partial}{\partial \phi}\right)^b
		\right]= 0,
	\label{wf}
\end{eqnarray}
where $F_{ab}$ is the antisymmetric electromagnetic field tensor. For the model 
adopted in this paper, $F_{ab}$ is given by the equation~(13) of \cite{li03b}. 
Then from equation~(\ref{wf}) we have $\Omega_{\rm F} = \Omega_\Psi$.
Equation~(\ref{wf}) means that the electric field measured by a pseudo-observer 
who rotates around the black hole with an angular velocity $\Omega_{\rm F}$ is 
vanishing.\footnote{We remind the reader that the electric field $E_a$ and the 
magnetic field $B_a$ measured by an observer of four-velocity $u^b$ are defined 
by the antisymmetric electromagnetic field tensor $F_{ab}$ through $E_a = 
F_{ab}u^b$ and $B_a = -\frac{1}{2}\epsilon_{abcd}u^b F^{cd}$, where 
$\epsilon_{abcd}$ is the totally antisymmetric tensor of the volume element 
associated with the spacetime metric $g_{ab}$.} In terms of the electric field 
and the magnetic field (who are 
perpendicular to each other) measured by an observer in the locally nonrotating 
frame, equation~(\ref{wf}) can be written as ${\bf E} = {\bf B}_{\rm p} \times {\bf 
v}_{\rm F}$ and ${\bf v}_{\rm F} = \chi^{-1} \left(\Omega_{\rm F} -\omega\right) 
{\bf m}$ [see eq.~[5.3] of \cite{mac82}], where $m^a = (\partial/\partial
\phi)^a$, ${\bf B}_{\rm p}$ is the poloidal component of the magnetic field.
Since the velocity of disk particles have both radial and azimuthal components 
while ${\bf v}_{\rm F}$---the linear velocity associated with the angular 
velocity $\Omega_{\rm F}$---has only an azimuthal component, generally 
$\Omega_{\rm F}$ is not equal to the disk angular velocity $\Omega_{\rm D}$ 
even though the magnetic field is frozen to the disk. Indeed, since $|{\bf E}|$ does 
not have to be smaller than or equal to $|{\bf B}_{\rm p}|$, the linear velocity 
${\bf v}_{\rm F}$ can be greater than the speed of light \citep{mac82}---hence
the name ``pseudo-observer'' used above. In addition, although the angular 
velocity of disk particles is usually a function of radius, for a stationary and 
axisymmetric configuration $\Omega_{\rm F}$ must be constant along magnetic 
field lines, otherwise magnetic field lines would wind themselves up in violation 
of stationarity. This is a relativistic extension of the isorotation law of 
\citet{fer37} and a mathematical proof can be found in \cite{car79}, \cite{mac82}, 
and \cite{li03b}.

The corresponding differential equation for $u^r$ can be derived from
equation~(\ref{funcf}) by using
\begin{eqnarray}
	\frac{\partial F}{\partial u^r} \frac{du^r}{dr}
		+ \frac{\partial F}{\partial r} = 0.
	\label{deq}
\end{eqnarray}
Equation~(\ref{deq}) contains two singularities: one is at the fast critical 
point ($r_{\rm f}$), the other is at the Alfv\'en critical point ($r_{\rm A}$). 
The Alfv\'en critical point, defined by $u_r u^r = c_{{\rm A}r} c_{\rm 
A}^{~\,r}$ where $c_{\rm A}^{~\,a}$ is the Alfv\'en velocity, turns out to be 
an apparent singularity, since it appears in both $\partial F/\partial u^r$ and 
$\partial F/\partial r$ so cancels out. However, the fast critical point, 
defined by $u_r u^r = c_{\rm A}^2/\left(1-c_{\rm A}^2\right)$ where $c_{\rm 
A}^2 = c_{{\rm A}a}c_{\rm A}^{~\,a}$, is a true singularity. For accretion 
solutions particles must leave the inner boundary of the disk 
``submagnetosonically'' but enter the horizon of the black hole 
``supermagnetosonically''. Thus, any physical solution must smoothly pass the 
fast critical point, which requires that $\partial F/\partial u^r = 0$ and 
$\partial F/\partial r = 0$ simultaneously at $r = r_{\rm f}$. Because of this 
restriction, among the four integral constants ($c_0$, $\Omega_\Psi$, $f_{\rm 
E}^\prime$, and $f_{\rm L}$) only three are independent: the other one has to 
be determined as a function of them \citep{li03b}.

The disk is assumed to be Keplerian (with a tiny inward radial velocity
superposed) and has an inner boundary at the marginally stable circular orbit: 
$r_{\rm in} = r_{\rm ms}$. Although we have argued that the angular velocity of 
magnetic field lines and the angular velocity of disk particles are two 
different concepts and in general they are not equal to each other, here 
following \citet{gam99} we assume that $\Omega_\Psi= \Omega_{\rm in}$, where
$\Omega_{\rm in}$ is the angular velocity of the disk at the inner boundary. 
This boundary condition implies that the magnetic field is locked to and 
corotates with the inner boundary of the disk, which is a good approximation
when the radial velocity is much smaller than the rotational velocity at the
inner boundary of the disk. The reasonableness of this 
assumption is supported by the asymptotic relation $f_{\rm E} \approx \Omega_\Psi 
f_{\rm L}$ as $c_0/r_{\rm g}\rightarrow \infty$, as we will see in the end of
Sec.~\ref{sec3}. Then, following the assumption $\Omega_\Psi= \Omega_{\rm in}$, 
the constant $f_{\rm E}^\prime$ is uniquely determined by the specific energy, 
the specific angular momentum, and the angular velocity of the disk at the inner 
boundary \citep{gam99,li03b}: $f_{\rm E}^\prime = - E_{\rm in} + \Omega_{\rm in} 
L_{\rm in}$, which is always negative.

Then, only one free parameter is left: $c_0$, and $f_{\rm L}$ will be determined 
as a function of $c_0$, $\Omega_\Psi$, and $f_{\rm E}^\prime$. From the 
definition of $c_0$, we have
\begin{eqnarray}
     \alpha_c &\equiv& \frac{c_0}{r_{\rm g}} \approx \frac{2 B_{\rm H}r_{\rm H}}
		    {\sqrt{\dot{M}_{\rm D}}} \nonumber\\
          &\approx& 5 \left[1+\left(1-\frac{a^2}{M^2}\right)^{1/2}\right]
              \left(\frac{B_{\rm H}}{10^4{\rm G}}\right)\left(\frac{M}{10^9 
              M_\odot}\right)\left(\frac{\dot{M}_{\rm D}}{10^{25}{\rm g/s}}
		    \right)^{-1/2}.
	\label{alphc}
\end{eqnarray}
where $r_{\rm H}$ is the radius of the black hole horizon, $B_{\rm H}$ is the
normal component of the magnetic field on the black hole horizon \citep{mac82}. 
The square of the dimensionless parameter $\alpha_c$, roughly
speaking, measures the ratio of the ``power'' of the magnetic connection to 
the ``power'' of accretion. As explained in the Introduction, for the model of 
magnetic connection between a black hole and an accretion disk, $\alpha_c$ can 
be $\gg 1$ in the limit of low accretion rate, although \citet{gam99} 
estimated that $\alpha_c \sim 1$ for the chaotic magnetic field in the 
transition region donated by accretion from a magnetized disk. Therefore, in 
our calculation we will treat $\alpha_c$ as a free parameter and allow it to 
vary from $10^{-1}$ to $10^4$. [The case of $\alpha_c \ll 1$ corresponds to 
the limit of high accretion rate, or equivalently weak radial magnetic field, 
as already studied by Li (\yearcite{li03a,li03b}).] 

Having specified the values of $\Omega_\Psi$ and $f_{\rm E}^\prime$, and 
the range of $c_0$, we are ready to solve the radial momentum equation $F
(r, u^r) =0$ for solutions that smoothly pass the fast critical point, and
determine the values of $f_{\rm L}$ and $f_{\rm E} = f_{\rm E}^\prime +
\Omega_\Psi f_{\rm L}$ for each solution, following the approach of 
\citet{li03b}. The results will be presented in the next section.

\section{Results}
\label{sec3}

The constant $f_{\rm L}$ can be solved from $F(r, u^r) = 0$. For $f_{\rm L}$ to
be real, the following condition must be satisfied
\begin{eqnarray}
     \left[\frac{r^2}{\Delta}\left(u^r\right)^2 + 1\right]\left[\chi^2 
		- \frac{A}{r^2}\left(\omega-\Omega_\Psi\right)^2\right] 
	     \le f_{\rm E}^{\prime 2}.
	\label{bound1}
\end{eqnarray}
At $r = r_{\rm f}$, $u^r$ can be expressed as a function of $r_{\rm f}$, then 
equation~(\ref{bound1}) becomes
\begin{eqnarray}
     \left[\chi^2- \frac{A}{r^2}\left(\omega-\Omega_\Psi\right)^2
	     \right]^3 \frac{c_0^4}{r^2} \leq 
          -\Delta\left[\chi^2- \frac{A}{r^2}\left(\omega -\Omega_\Psi
		\right)^2 -f_{\rm E}^{\prime 2}\right], \hspace{0.6cm}
	     r = r_{\rm f}.
     \label{bound2}
\end{eqnarray}
Equation~(\ref{bound2}) restricts the physically allowed region in the
$(r_{\rm f}, c_0)$-space, while equation~(\ref{bound1}) restricts the physically 
allowed region in the $(r,u^r)$-phase space.

\begin{figure}
\begin{center}
\FigureFile(12cm,){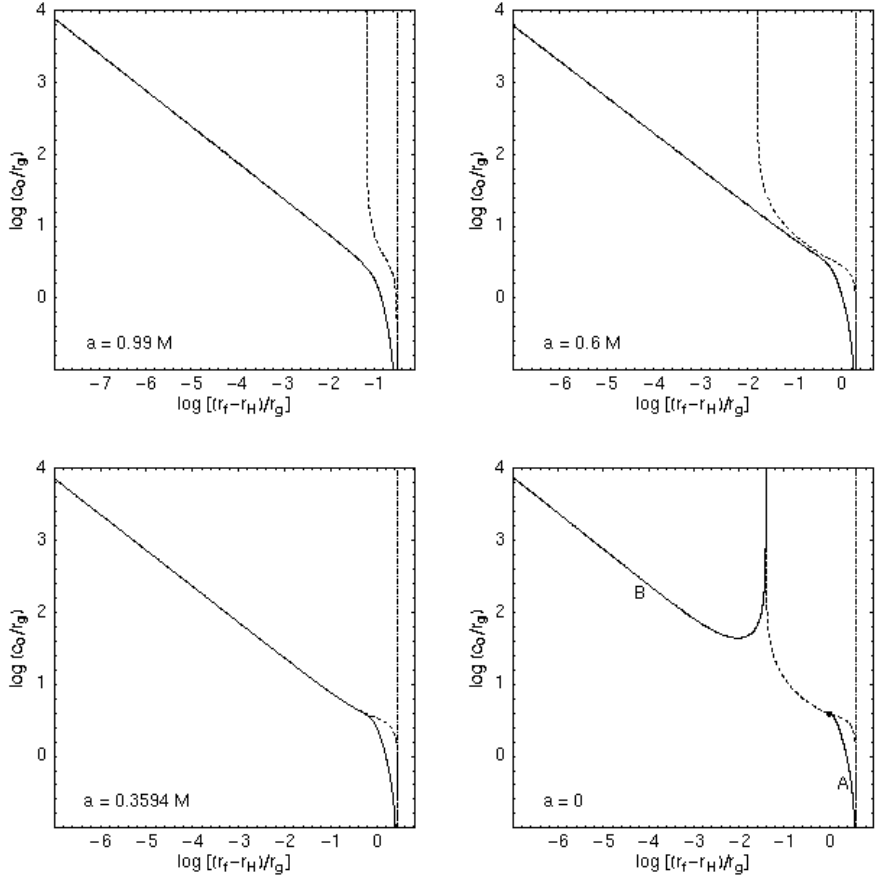}
\end{center}
\caption{Solutions for the radius at the fast critical point, $r_{\rm f}$
(solid curves). Each panel corresponds to a different spinning state of 
the black hole, as indicated by the value of $a$. The outer boundary of 
the flow is at the marginally stable circular orbit, as indicated by the 
vertical dotted line. The dashed curve represents the boundary for physical 
solutions: beyond the dashed curve physical solutions do not exist (see 
the text).
\label{fig1}}
\end{figure}

With $f_{\rm L}$ expressed in terms of $r_{\rm f}$ and $u_{\rm f}^r \equiv u^r
(r_{\rm f})$, we can solve for $r_{\rm f}$ and $u_{\rm f}^r$ from the condition
$\partial F/\partial r = 0 = \partial F/\partial u^r$ at $r = r_{\rm f}$ 
\citep{li03b}. The solutions for $r_{\rm f}$ as a function of $\alpha_c = 
c_0/r_{\rm g}$ are shown in Fig.~\ref{fig1}. Each panel corresponds to a 
different spinning state of the black hole, as indicated by the value of $a$. 
The vertical dotted line shows the position of the marginally stable circular 
orbit, i.e., the inner boundary of the disk. The dashed curve represents the 
boundary for physical solutions, beyond which equation~(\ref{bound2}) is violated 
so physical solutions do not exist. 

From Fig.~\ref{fig1} we see that, the solutions for $r_{\rm f}$ can be 
divided into two classes according to the different topology they have, one
class corresponds to $a > a_{\rm cr}\equiv 0.3594 M$, the other corresponds 
to $0\leq a < a_{\rm cr}$. (Throughout the paper we assume that $0\leq a/M < 
1$.) The critical spin $a_{\rm cr}$ corresponds to a state that the Keplerian 
disk angular velocity at the marginally stable circular orbit is equal to the 
spinning angular velocity of the black hole, i.e. the disk inner boundary 
corotates with the black hole horizon \citep{li00a,li02a}. When $a > a_{\rm 
cr}$ (then the black hole rotates faster than the inner boundary of the disk 
and any point beyond that), there is only one branch of solutions for $r_{\rm 
f}$, which is separated from the boundary for physical solutions. As $a
\rightarrow a_{\rm cr}$ from above, the boundary for physical solutions merges 
to the solutions for $r_{\rm f}$, when $c_0/r_{\rm g}$ is large (see the 
bottom-left panel). When $0\leq a < a_{\rm cr}$ (then the black hole rotates 
slower than the inner boundary of the disk), there are two disconnected branches 
(A and B) of solutions for $r_{\rm f}$. Branch A ends at the boundary for 
physical solutions as $c_0/r_{\rm g}$ increases (see the dark point in the 
bottom-right panel). As $a$ increases from $0$ to $a_{\rm cr}$, the branch B 
and the end point of branch A move toward the up-left direction, until they 
disappear when $a=a_{\rm cr}$ is reached.

For the case of $a = 0$, and generally for any $0\leq a < a_{\rm cr}$, the 
solution for $r_{\rm f}$ on branch B does not correspond to a physical 
solution representing a flow that starts from $r_{\rm in}$ and ends at $r_{\rm 
H}$, see Appendix~\ref{app1}. Therefore, for any $0\leq a < a_{\rm cr}$, there 
is a maximum value of $\alpha_c$, beyond which stationary and axisymmetric 
solutions do not exist. For the case of $a =0$ (the Schwarzschild case), we 
have $\alpha_{c,\max} \approx 3.88$, corresponding to $r_{\rm f} \approx 3 
r_{\rm g}$ (the dark dot in the bottom-right panel of Fig.~\ref{fig1}).

Note, for all cases, $r_{\rm f}$ approaches $r_{\rm in} = r_{\rm ms}$ as
$c_0^2/r_{\rm g}^2 \rightarrow 0$, confirming the earlier results of Li
(\yearcite{li03a,li03b})---although $\Omega_\Psi\neq 0$ here.

With the solutions for $r_{\rm f}$ found, we can calculate $f_{\rm L}$, then
$f_{\rm E}$. Then, we can solve for the global solutions that start from the
inner boundary of the disk, smoothly pass the fast critical point, and end 
at the horizon of the black hole. Examples of such solutions are presented
in Appendix~\ref{app1}. Those examples confirm the conclusions that global 
solutions exist, except for parameters on the branch B when $0\le a <a_{\rm 
cr}$. In this section, we present some integral results, since those integral 
results are more closely related to observational effects. 

\begin{figure}
\begin{center}
\FigureFile(12cm,){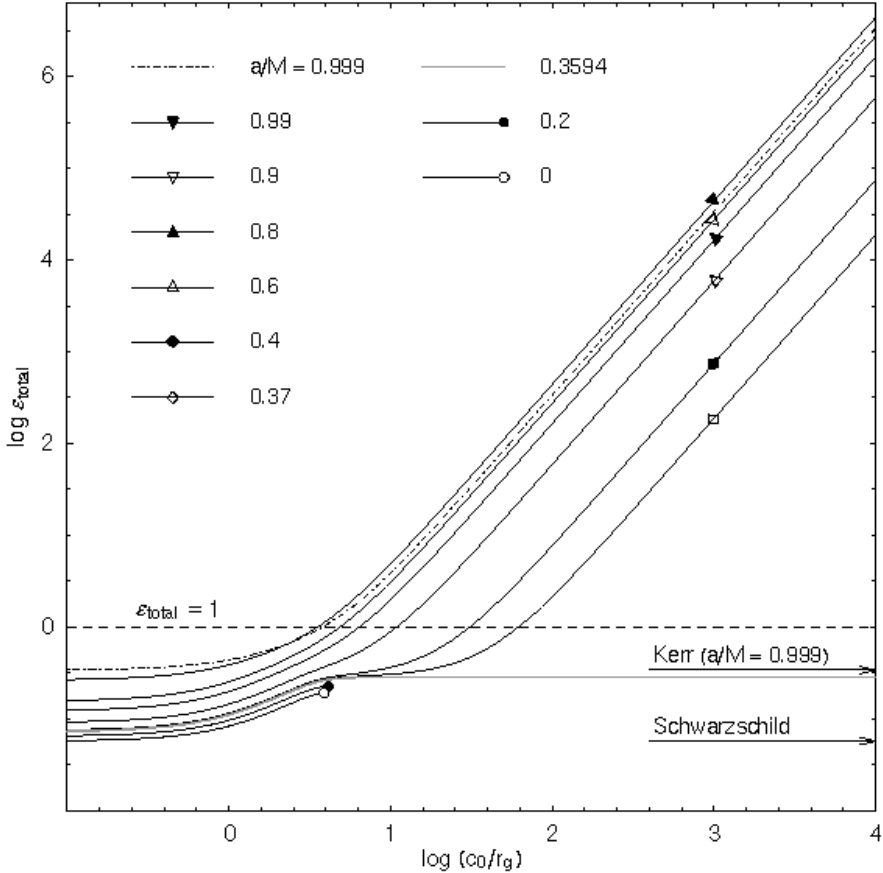}
\end{center}
\caption{The total radiation efficiency of the system, $\varepsilon_{\rm
total} = 1 + f_{\rm E}$, as a function of $\alpha_c = c_0/r_{\rm g}$. 
Different lines correspond to different spinning states of the black hole,
as labeled. For reference, the following typical values for disk radiation 
efficiencies are indicated: $0.057$, corresponding to a standard Keplerian
disk around a Schwarzschild black hole; $0.34$, a standard Keplerian disk
around a Kerr black hole with $a = 0.999 M$; and $1$, beyond which some of
radiated energy must be tapped from the black hole. 
\label{fig2}}
\end{figure}

First, the total radiation efficiency of the system, defined by the ratio of 
the total power of the disk to the mass accretion rate,\footnote{The transition 
region is not likely to be able to radiate much of energy since significant 
change in the specific energy of particles inside the marginally stable orbit 
necessarily implies that the non-gravitational force is important then the 
material in the transition region must be geometrically thick so have low 
radiative efficiency.} can be calculated by (see, e.g., \cite{gam99})
\begin{eqnarray}
     \varepsilon_{\rm total} = 1 + f_{\rm E}.
	\label{etotal}
\end{eqnarray}
The results for $\varepsilon_{\rm total}$ are shown in Fig.~\ref{fig2} as a 
function of $\alpha_c$. For the cases of $a = 0$ and $a = 0.2 M$, the curves
end at $\alpha_{c,\max}=3.88$ and $\alpha_{c,\max}=4.10$, respectively.

Figure~\ref{fig2} shows that, although as $c_0^2/r_{\rm g}^2 \rightarrow 0$ all
efficiencies approach the corresponding values for a standard Keplerian disk,
as $c_0^2/r_{\rm g}^2$ goes up the disk radiation efficiency is increased by 
the magnetic connection. Furthermore, a disk around a Kerr black hole with 
$a>a_{\rm cr}$ behaves very differently from that around a Kerr black hole 
with $0\leq a< a_{\rm cr}$. For $a>a_{\rm cr}$, as $c_0/r_{\rm g}$ increases 
the total efficiency $\varepsilon_{\rm total}$ keeps increasing without limit. 
For sufficiently large $c_0/r_{\rm g}$, $\varepsilon_{\rm total}$ can be 
greater than unity, then a part of the energy radiated by the disk must be 
extracted from the spin 
energy of the black hole. Note, as $a$ goes from $a_{\rm cr}$ to $a = 0.99M$, 
$\varepsilon_{\rm total}$ keeps going up; but after that, as $a$ goes from
$0.99M$ to $a = 0.999 M$, $\varepsilon_{\rm total}$ goes down for large
$c_0/r_{\rm g}$. This is caused by the fact that as $a\rightarrow
M$, the spinning angular velocity of the black hole approaches the angular
velocity at the inner boundary of the disk, so if $a$ is very close to $M$
the power of energy extracted from the black hole decreases with increasing
$a/M$ (see, e.g., Li \yearcite{li00a,li02a}). 

For $0\leq a< a_{\rm cr}$, as $c_0/r_{\rm g}$ goes up from small values, 
$\varepsilon_{\rm total}$ increases until a maximum of $\varepsilon_{\rm 
total}$ is reached at $\alpha_c = \alpha_{c,\max}$. For $c_0/r_{\rm g} >
\alpha_{c,\max}$, stationary and axisymmetric solutions do not exist. We 
speculate
that, for $0\leq a< a_{\rm cr}$ and large $c_0/r_{\rm g}$, the inner part 
of the disk will be disrupted by the magnetic field, resulting that the 
inner boundary of the disk moves outward. (More implications will be 
discussed near the end of this section.) Even for $c_0/r_{\rm g} < \alpha_{c,
\max}$, the magnetic field can still be dynamically important and leads to
to increase in the total radiation efficiency of the disk. For example, for
the case of a Schwarzschild black hole ($a =0$), the total radiation 
efficiency at $c_0/r_{\rm g} = \alpha_{c,\max}$ is $\approx 0.19$, $3.4$
times higher than the corresponding value ($\approx 0.057$) for a standard 
Keplerian disk with no magnetic connection.

For the critical case of $a = a_{\rm cr}$, the radiation efficiency increases
as $c_0/r_{\rm g}$ goes up, but quickly approaches a constant for $c_0/r_{\rm 
g}>10$. This can be shown analytically. From Fig.~\ref{fig1}, the bottom-left
panel, we see that for $a = a_{\rm cr}$ and large $c_0/r_{\rm g}$, the 
boundary for physical solutions merges to the solutions for $r_{\rm f}$. Then,
from equation~(\ref{funcf}) we have 
\begin{eqnarray}
     f_{\rm L} = - \left[\frac{\frac{A}{r^2}\left(\omega-\Omega_\Psi\right)}
	     {\chi^2 - \frac{A}{r^2}\left(\omega-\Omega_\Psi\right)^2}
		\right]_{r = r_{\rm f}}f_{\rm E}^\prime.
	\label{fl}
\end{eqnarray}
As $c_0/r_{\rm g}\rightarrow\infty$, we have $r_{\rm f}\rightarrow r_{\rm H}$
and $\omega(r = r_{\rm f}) \rightarrow \Omega_{\rm H} = \Omega_\Psi$ (for
$a = a_{\rm cr}$). Then, both the numerator and denominator on the right-hand
of equation~(\ref{fl}) approach zero as $c_0/r_{\rm g}\rightarrow\infty$, the 
limiting value in the brackets can be evaluated by L'Hospital's rule. The
results for $f_{\rm L}$ and $f_{\rm E}$ are then
\begin{eqnarray}
     f_{\rm L} \approx 1.1687 r_{\rm g} f_{\rm E}^\prime \approx -0.7488\,
	     r_{\rm g}, \hspace{0.6cm}
	f_{\rm E} \approx -0.7103,
	\label{fl2}
\end{eqnarray}
which are independent of $c_0/r_{\rm g}$ provided that $c_0^2/r_{\rm g}^2 \gg
1$. The corresponding limiting radiation efficiency is $\varepsilon_{\rm total}
(a = a_{\rm cr},c_0^2/r_{\rm g}^2 \gg 1) \approx 0.29$, four times higher than 
that for a standard disk around a Kerr black hole of $a = a_{\rm cr}$.
 
\begin{figure}
\begin{center}
\FigureFile(12cm,){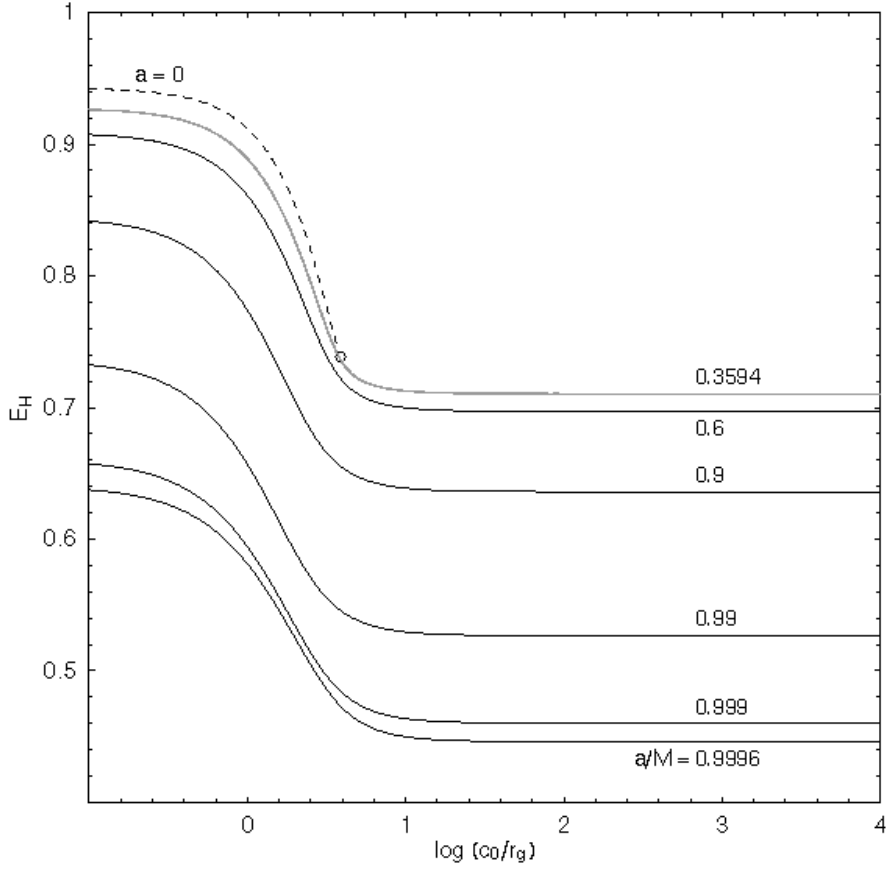}
\end{center}
\caption{The specific energy of disk particles as they reach the horizon of 
the black hole, as a function of $c_0/r_{\rm g}$. Different lines correspond
to different spinning states of the black hole, as labeled. As $c_0^2/r_{\rm 
g}^2\rightarrow 0$, $E_{\rm H}$ approaches the specific energy on the 
marginally stable circular orbit.
\label{fig3}}
\end{figure}

The inequality $\varepsilon_{\rm total} >1$ is a sufficient condition for a 
part of the energy radiated by the disk being extracted from the spin energy of 
the black hole, since when $\varepsilon_{\rm total} >1$ the net energy flux 
$F_{\rm E}$ must be positive. This can also be explained by the fact that the 
maximum amount of energy that
can be extracted from the disk material is equal to the rest mass of the disk.
However, $\varepsilon_{\rm total} >1$ is not a necessary condition,
which means that a part of energy can still be extracted from the black hole
if $\varepsilon_{\rm total} \leq 1$. Indeed, for all models calculated in
Fig.~\ref{fig2}, the specific energy of disk particles are always finite and
positive as they reach the horizon of the black hole, see Fig.~\ref{fig3}.
For large $c_0/r_{\rm g}$, the specific energy of particles on the horizon
of the black hole, $E_{\rm H}$, is smaller than the specific energy of
particles on the marginally stable circular orbit by a significant fraction, 
implying that the magnetic field in the transition region is dynamically 
important and keeps removing energy and angular momentum from particles. 
However, $E_{\rm H}$
never gets to zero no matter how large $c_0/r_{\rm g}$ is. In fact, 
Fig.~\ref{fig3} shows that, for $a\geq a_{\rm cr}$, $E_{\rm H}$ approaches
positive constants as $c_0/r_{\rm g}\rightarrow\infty$. This makes the
approach of extracting energy from a Kerr black hole with magnetic fields
distinctly different from the Penrose mechanism \citep{pen69} (c.f. 
\cite{hir92,wil03}): In the former case negative electromagnetic energy flows 
into the horizon of the black hole but the energy of particles remains positive, 
while in the latter the energy of the particle that falls into the black hole 
has to be negative.

\begin{figure}
\begin{center}
\FigureFile(12cm,){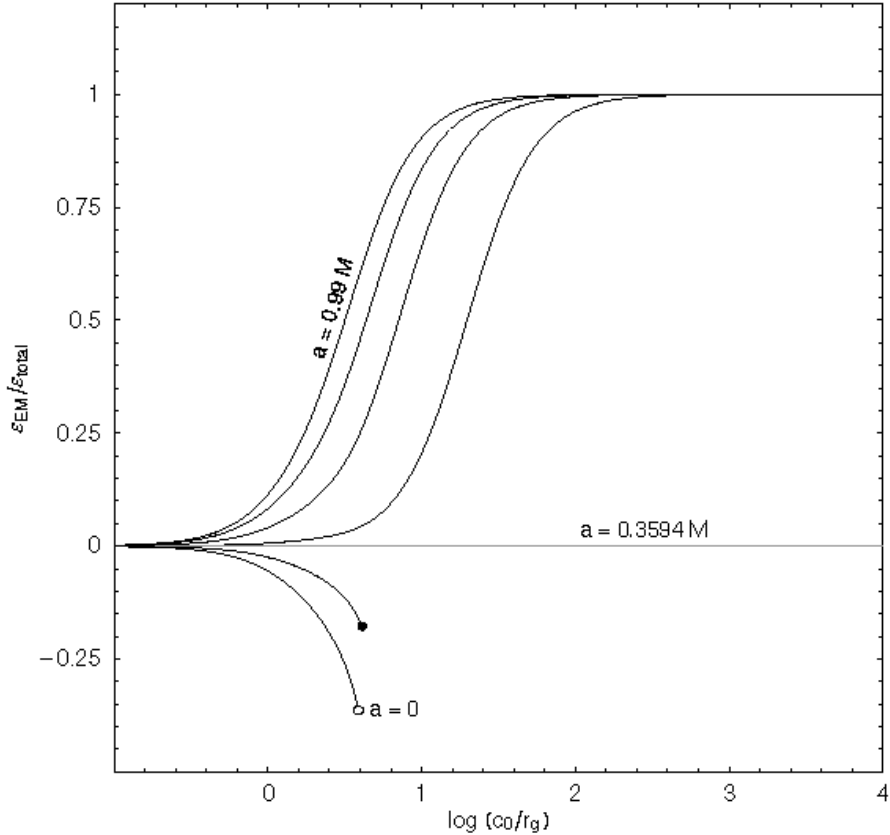}
\end{center}
\caption{Fraction of the electromagnetic energy (Poynting flux) contribution 
to the total radiation efficiency. When $\varepsilon_{\rm EM}/\varepsilon_{\rm
total} >0$, $\varepsilon_{\rm EM}/\varepsilon_{\rm total}$ of the total 
radiated energy is extracted from the spin energy of the black hole in the 
form of Poynting flux. When $\varepsilon_{\rm EM}/\varepsilon_{\rm total} <
0$, part of the mechanical energy of particles is converted to the energy 
of electromagnetic fields and falls into the black hole. Different lines 
correspond to different spinning states of the black hole: $a/M = 0$, 
$0.2$, $0.3594$, $0.4$, $0.6$, $0.8$, and $0.99$ (bottom-up).
\label{fig4}}
\end{figure}

The total radiation efficiency, $\varepsilon_{\rm total}$, can be written as
a sum of two distinct contributions: the contribution from the mechanical
energy (including rest-mass energy) of particles, $\varepsilon_{\rm mech} = 
1 - E_{\rm H}$; and the contribution from the electromagnetic energy (i.e.,
the Poynting flux), $\varepsilon_{\rm EM} = \varepsilon_{\rm total} - 
\varepsilon_{\rm mech} = f_{\rm E} + E_{\rm H}$. For the model studied in 
this paper, $\varepsilon_{\rm mech}$ is always positive, while 
$\varepsilon_{\rm EM}$ can take either sign. When $\varepsilon_{\rm EM}> 0$,
electromagnetic energy flows out of the black hole horizon, part of the
radiated energy is extracted from the spin energy of the black hole. When 
$\varepsilon_{\rm EM} <0$, part of the mechanical energy of disk particles 
is converted into electromagnetic energy and flows into the black hole 
horizon, the rest is transported to the disk then radiated away, no energy is 
extracted from the 
black hole. Thus, when $\varepsilon_{\rm EM} \leq 0$, all the energy radiated 
by the disk eventually comes from the gravitational binding energy of disk 
particles. The fraction $\varepsilon_{\rm EM}/\varepsilon_{\rm total}$ is
calculated and shown in Fig.~\ref{fig4}, where different curves correspond 
to different spinning states of the black hole. We see that, extraction of
energy from black holes is possible only if $a> a_{\rm cr} = 0.3594 M$.
For $a< a_{\rm cr}$, a fraction of the mechanical energy of particles is
converted to the electromagnetic energy which flows into the horizon of the
black hole. In the critical case $a= a_{\rm cr}$, no electromagnetic energy 
flows into or out of the black hole, the change in the mechanical energy of 
particles is all eventually radiated away by the disk. We also see that, for 
$a> a_{\rm cr}$, extraction of energy from black holes happens even when 
$\varepsilon_{\rm total} <1$. For example, when $a = 0.99 M$ and $c_0 = 
2 r_{\rm g}$, we have $\varepsilon_{\rm total} \approx 0.59$ and 
$\varepsilon_{\rm EM}/\varepsilon_{\rm total} \approx 0.30$, so about $30\%$
of the total radiation comes from the spinning energy of the black hole.

\begin{figure}
\vspace{0.07cm}
\begin{center}
\FigureFile(12cm,){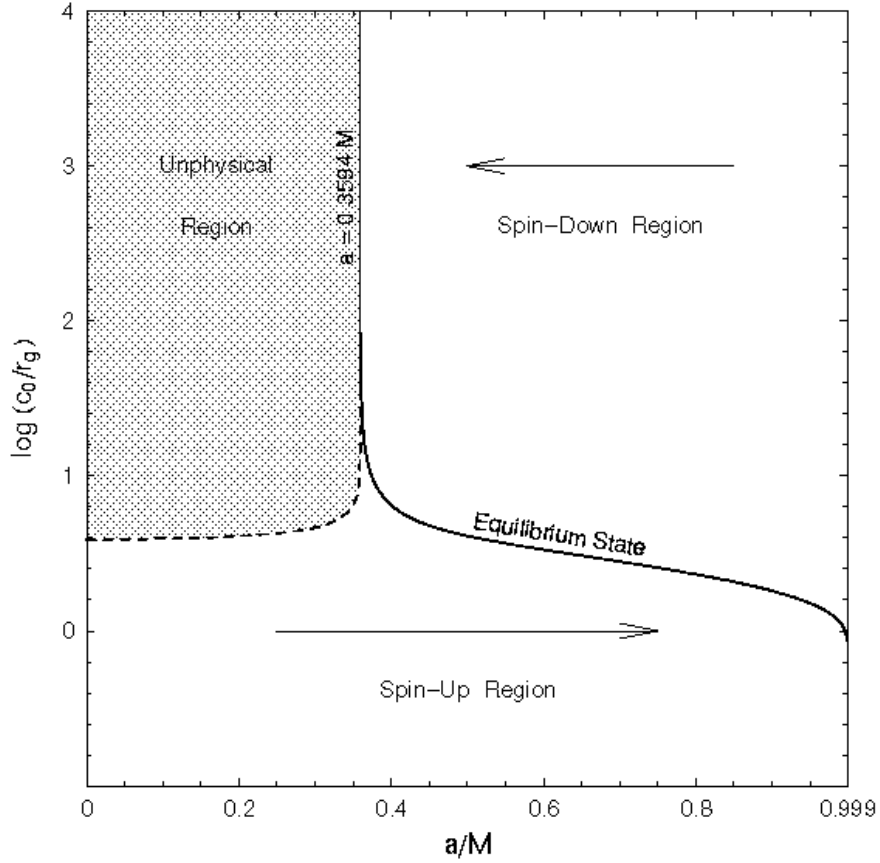}
\end{center}
\caption{The $(a,c_0)$-space is divided into three regions: the spin-down 
region, the spin-up region, and the region where physical solutions do not 
exist (shaded region). If a black hole-accretion disk system is in the 
spin-down region, the black hole will be spun down (i.e., its spin $a/M$ 
decreases with time) by the joint action of magnetic connection and 
accretion (as indicated by the leftward arrow). If a black hole-accretion 
disk system is in the spin-up region, the black hole will be spun up by the 
magnetic connection and accretion (as indicated by the rightward arrow). 
The two regions are separated by the equilibrium state (the solid curve), 
in which the spin of the black hole does not change with time. The curve
for the equilibrium state merges to the line of $a = 0.3594M$, as $c_0/
r_{\rm g}\rightarrow\infty$. (The inner boundary of the disk is at the
marginally stable circular orbit.)
\label{fig5}}
\end{figure}

It is interesting to check the evolution of the black hole spin, $a/M$, under 
the joint action of magnetic connection and accretion. The evolution direction
of the black hole spin is determined by the sign of $f_{\rm L} - 2a f_{\rm 
E}$: If $f_{\rm L} - 2a f_{\rm E} >0$, the black hole spins down (i.e., 
$a/M$ decreases with time); if $f_{\rm L} - 2a f_{\rm E} <0$, the black hole
spins up (i.e., $a/M$ increases with time). If $f_{\rm L} - 2a f_{\rm E}$
happens to be zero, then the spin of the black hole is in an equilibrium state 
and does not change with time \citep{gam99}. In Fig.~\ref{fig5} we show the 
evolution direction of the black hole spin in the $(a,c_0)$-space. We see 
that, for all models with $a < a_{\rm cr}$, the black hole spins up. While for 
models with $a > a_{\rm cr}$, the evolution direction of the black hole spin 
depends on the value of $c_0/r_{\rm g}$. For a given $a> a_{\rm cr}$, the 
black hole spins down/up if $c_0/r_{\rm g}$ is greater/smaller than some 
critical value. As an example, for $a/M = 0.99$, the critical value of $c_0/
r_{\rm g}$ is $\approx 1.12$. For $c_0/r_{\rm g}\gg 1$, the equilibrium state
(corresponding to $f_{\rm L} - 2a f_{\rm E} =0$) approaches $a_{\rm cr}/M = 
0.3594$ from above. 

\begin{figure}
\begin{center}
\FigureFile(12cm,){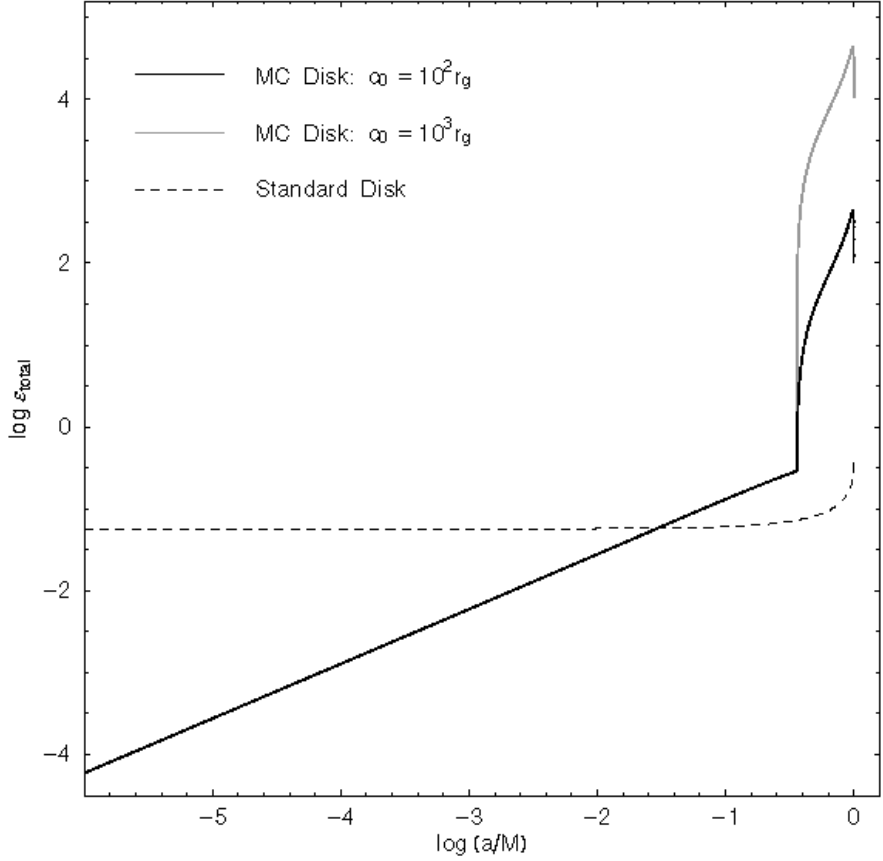}
\end{center}
\caption{The total radiation efficiency of a black hole-accretion disk system 
with magnetic connection (MC), where the inner boundary radius of the Keplerian 
disk is given by equation~(\ref{rin}). Two models are calculated: one with $c_0/
r_{\rm g} = 10^2$, the other with $c_0/r_{\rm g} = 10^3$. The curves break at 
$(a/M,\varepsilon_{\rm total}) = (0.3594,0.2897)$. For comparison, the
radiation efficiency corresponding to a standard Keplerian disk is shown
with the dashed line.
\label{fig6}}
\end{figure}

It is worth to make some further comments on the implications of our results on 
black hole-accretion disk systems with $0\le a< a_{\rm cr}$ and large $c_0/r_{\rm 
g}$. As we have shown, for sufficiently large $c_0/r_{\rm g}$ and $0\le a< 
a_{\rm cr}$, stationary and axisymmetric solutions do not exist. This 
implies that when $c_0/r_{\rm g}$ is large, the inner part of a Keplerian 
disk around a Kerr black hole with $0\le a< a_{\rm cr}$ will be 
disrupted by the strong magnetic field connecting the black hole to the disk.
As a result, the inner boundary of the disk moves out, until a corotation
radius is reached where the Keplerian angular velocity of the disk is 
equal to the spinning angular velocity of the black hole. Then, a new 
stationary equilibrium state is established, with a Keplerian disk truncated
inwardly at the corotation radius
\begin{eqnarray}
     r_{\rm c} = M \left(\frac{a}{M}\right)^{-2/3} \left[
	     1 + \sqrt{1- \left(\frac{a}{M}\right)^2}\right]^{4/3},
	\label{rc}
\end{eqnarray}
for $0\leq a < a_{\rm cr}$. This motivates us to propose the following model 
for a black hole-accretion disk system with large $c_0/r_{\rm g}$: the inner
boundary of the geometrically thin Keplerian disk is given by
\begin{eqnarray}
     r_{\rm in} = \left\{\begin{array}{ll}
	          r_{\rm ms}, \hspace{0.5cm}~ & a>a_{\rm cr}; \\
			r_{\rm c},  \hspace{0.5cm}~ & 0\leq a < a_{\rm cr}.
			\end{array}
	          \right.
	\label{rin}
\end{eqnarray}
For $r< r_{\rm in}$, the magnetic field is dynamically important so the
material is geometrically thick. The magnetic field exerts a torque at the 
inner boundary of the disk, which pumps angular momentum and energy into
the Keplerian disk region. The total radiation efficiency of such a disk is 
shown in Fig.~\ref{fig6} as a function of the black hole spin. 

Figure~\ref{fig6} shows that, the total radiation efficiency of a disk with 
a strong magnetic connection dramatically differs from that of a standard 
Keplerian accretion disk. For $a>a_{\rm cr}$, the spin energy of the black 
hole is efficiently extracted by the magnetic field and fed into the disk,
increasing the disk radiation efficiency dramatically. (As $a/M$ approaches
unity, the radiation efficiency drops since when $a/M = 1$ the angular
velocity at the inner boundary of the disk is equal to the spin angular
velocity of the black hole.) While for $0\leq a<a_{\rm cr}$, the radiation 
efficiency is independent of the value of $c_0/r_{\rm g}$, provided that $c_0
/r_{\rm g}$ is large. For $0.03 < a/M < 0.3594$, the radiation efficiency is 
still larger than that of a standard Keplerian disk despite the fact the 
inner boundary of the disk moves out. This is caused by the nonzero torque at 
the disk inner boundary produced by the magnetic field, which transports 
energy from the transition region to the disk. For $0\leq a/M <0.03$, the 
effect of the growth in the disk inner boundary radius becomes more important, 
resulting that the disk radiation efficiency becomes smaller than that 
predicted by the standard disk model. As $a\rightarrow 0$, the disk radiation 
efficiency drops to zero (compare it to the corresponding efficiency $\approx 
0.057$ for a standard Keplerian disk around a Schwarzschild black hole).

For $0< a/M \ll 1$, we have $r_{\rm in} = r_{\rm c} \approx M (a/4M)^{-2/3} 
\gg r_{\rm g}$. Correspondingly, we have $f_{\rm E} \approx f_{\rm E}^\prime
\approx -1 + 3M/2 r_{\rm c}$, so we have $\varepsilon_{\rm total} = 1 +f_{\rm 
E} \approx 3M/2r_{\rm c} \approx 0.6\, (a/M)^{2/3}$. Thus, for a system 
with an extremely slow rotating black hole, the radiation efficiency of the 
accretion disk can be extremely low.

Finally, when $a>a_{\rm cr}$ and in the limit $c_0/r_{\rm g}\rightarrow\infty$, 
by equation~(57) of \cite{li03b} (and the fact that $E_{\rm H}$ and $L_{\rm H}$
remain finite, see Fig.~\ref{fig3}) we have $\varepsilon_{\rm total} \approx 
f_{\rm E} \approx \Omega_\Psi f_{\rm L}$, and the power of the black hole
\begin{eqnarray}
     P_{\rm H} \approx - F_{\rm m} f_{\rm E} \approx
	     \frac{2M}{r_{\rm H}}\, C_0^2\, \Omega_{\rm in} \left(\Omega_{\rm 
		H} - \Omega_{\rm in}\right),
	\label{ph}
\end{eqnarray}
where we have used $\Omega_\Psi = \Omega_{\rm in}$. From the definition of
$C_0$ \citep{li03b},  we have $C_0 = 2 B_{\rm H} M r_{\rm H}$. Equation~(\ref{ph}) 
can be used to estimate the power of a disk magnetically coupled to a fast 
rotating Kerr black hole.

\section{Discussion and Conclusions}
\label{sec4}

We have studied the energetics of a black hole-accretion disk system with a 
large-scale magnetic field connecting the disk to the black hole through the
transition region. The model that we have adopted is an extension of those used
by \citet{gam99} and \citet{li03b}. In \cite{gam99} (as well as in \cite{kro99}), 
the magnetic field is assumed to have small scales, be advected into the 
transition region from the disk by accretion, and couple the material
in the transition region to the disk. While in our model, we assume that the
magnetic field has a large scale, directly connect the disk to the black hole.
The role of accretion is only to provide a dilute plasma gas in the transition
region so that an electric current can exist there. In \cite{li03b}, the
large-scale magnetic field is assumed to extend from the black hole horizon
to infinity and have a zero angular velocity although the disk has a nonzero
angular velocity and the magnetic field is frozen to the disk (see 
Sec.~\ref{sec2} for the difference between the two angular velocities). While
in our model, the large-scale magnetic field is assumed to terminate at the 
inner boundary of a thin Keplerian disk, beyond which the magnetic field is 
chaotic, has small scales, and is dynamically unimportant. In addition, the 
magnetic field is assumed to corotate with the inner boundary of the disk.

We have focused our calculations to a small neighborhood of the equatorial plane,
although the material in the transition region must be geometrically thick when
the dynamical effects of magnetic fields are important. It
is assumed that the system is in a stationary and axisymmetric state, and in
the small neighborhood, where $\cos^2\theta\ll 1$, both the magnetic field and 
the velocity field have only radial and azimuthal components (i.e., $B^\theta = 
u^\theta =0$). This model is similar to the cylindrical model that is often 
used for accretion disks, where the cylindrical coordinates are used and every
quantity is independent of the vertical coordinate. Here we use the spherical
coordinates and assume that every quantity is independent of $\theta$---which
is more appropriate for an accretion flow near the central black hole. The
self-consistency of this model has been proved by \citet{li03b}, where it was
shown that to the first order of $\cos\theta$ the dynamical equilibrium in the
$\theta$-direction is guaranteed. The extension to the regions well above and
below the equatorial plane has also been discussed by \citet{li03b}.

With our simplified model, we find that in the limit of low mass accretion rate, 
a system with a fast rotating black hole and a system with a slow rotating 
black hole behave very differently. For a black hole with $a > a_{\rm cr} = 
0.3594 M$, the spin energy of the black hole is efficiently extracted by the 
magnetic field and transported to the disk region, increasing the radiation
efficiency of the disk dramatically. Indeed, in such a case the total radiation 
efficiency of the disk is unbounded from above. For a black hole with $0\leq a 
<a_{\rm cr}$, stationary solutions do not exist if we assume the inner boundary
of the disk is at the marginally stable orbit, which leads us to speculate that 
in such a case the inner region of the disk is disrupted by the magnetic field 
and the inner boundary of the disk moves outward until a corotation radius 
(defined by $\Omega_{\rm D} = \Omega_{\rm H}$) is reached. When $0\leq a/M \ll 
1$, the disk---which has a very large radius at the inner boundary---has an 
extremely low radiation efficiency: $\varepsilon_{\rm total} \approx 0.6 
(a/M)^{2/3} \ll 1$. The above results are in great contrast to those for a 
standard geometrically thin Keplerian disk, whose radiation efficiency is 
always in the range $0.06$ -- $0.4$.

The disruption of the inner region of a disk magnetically coupled to a slow
rotating black hole can be interpreted by the following fact: When the black hole
rotates slower than the inner boundary of the disk, the black hole exerts a
negative torque to the inner boundary of the disk, but a Keplerian disk cannot
sustain a negative torque \citep{li00a,li02a}. The general consequences of the 
magnetic interaction between a black hole and an accretion disk have already
been explored in the literature (see, e.g., \cite{li00a,li02a}), while the 
calculations presented in this paper confirm those results with mathematics.

Note, the condition for the validity of the solutions is $\cos^2\theta\ll 1$, 
not $|\cos\theta|\ll 1$. For $71.6^\circ < \theta < 108.4^\circ$, we have 
$\cos^2\theta < 0.1$ where we expect the solutions are not bad. This corresponds 
to an opening angle of $37^\circ$ around the equatorial plane, which is not a 
small angle. Therefore, although our calculations are restricted in a thin slab 
close to the equatorial plane, we expect that the solutions are typical for the 
whole transition region. This conclusion is reasonable when the following fact 
is also considered: the accretion flow has highest density on the equatorial 
plane so the transportation of mass, energy, and angular momentum should dominantly 
happen in the region near the equatorial plane.

Hence, according to both the general results in the literature and the detailed
calculations in this paper, an accretion disk magnetically coupled to a black 
hole can have a 
radiation efficiency in a broad range: from an extremely low value $\ll 1$ to 
an extremely high value $\gg 1$, depending on the spin of the black hole, the 
strength of the magnetic field, and the mass accretion rate. This is of great 
interest since observations show that galactic nuclei at high and low redshifts 
have extremely divergent distribution in luminosities, from the extremely bright 
active galactic nuclei (AGN) at high redshift to the much less active---sometimes 
not active at all---nearby galactic nuclei, which is certainly not because of a 
lack of supermassive black holes in nearby galactic nuclei \citep{nar02}. In 
order to interpret the extremely low luminosities of nearby galactic nuclei, 
people were led to consider models of advection- or convection-dominated flow 
(ADAF/CDAF), and advection-dominated inflow outflow solution (ADIOS) 
(for reviews see \cite{bla99,nar02}). Our results on the system with a slow 
rotating 
black hole suggests a new interpretation for the low luminosities in nearby 
galactic nuclei: nearby galactic nuclei may have extremely slow rotating 
supermassive black holes and the inner region of the disks may have been 
disrupted by a strong magnetic field connecting the disk to the black 
hole.\footnote{An accretion disk around a fast rotating black hole, containing a 
large-scale magnetic field and having a high accretion rate, may also have an 
extremely low radiation efficiency and luminosity if the large-scale magnetic 
field extends to infinity \citep{li03b}.}

Apparently, the model presented in this paper has been highly simplified. In
a real situation, the magnetic field should have a more complex structure:
The foot points of magnetic field lines should distribute over the surface of 
the disk rather than all going through the inner boundary of the disk, and 
some of the magnetic field lines passing through the black hole may extend to 
infinity directly (see, e.g., \cite{nit91,bla02,uzd03}). The field lines 
connecting the black hole to the surface of the disk lead to additional 
contribution to the transportation of angular momentum and energy between the
black hole and the disk \citep{li02a,li02b,wan03}. The field lines connecting
the black hole directly to the plasma at infinity extract energy and angular
momentum from the black hole and transport them to remote plasma through the 
Blandford-Znajek mechanism, so will affect the evolution of the central black 
hole \citep{par88,par90,lu96,wan02}. Clearly, all these effects will affect 
some of the calculations in this paper, including the total radiation efficiency 
of the disk and the evolution of the spin of the black hole. Detailed 
calculations including the above effects are beyond the scope of the current
paper. However, we should emphasize that the main results in this paper are 
more qualitative rather than qualitative, they are accurate only if the effects
mentioned above are weak. The readers should keep this point in their mind to
correctly understand and apply these results.  

Finally, let us briefly comment on under what conditions a large-scale 
magnetic field connecting a disk and a black hole through the transition 
region can be formed. It appears to us that there are three possible ways: 
(1) Inherited from the progenitor of the black hole and the disk. For
example, merger of a magnetized neutron star with a black hole will naturally 
lead to the formation of a disk with a large-scale magnetic field. The 
turbulent motion [or, the magneto-rotational instability \citep{bal91,bal98}] 
in the disk region then disrupts the initially large-scale magnetic field, 
transform it into highly tangled small-scale magnetic fields. The magnetic
field in the transition region, on the other hand, may remain being ordered
and having large correlation scales. (2) Large-scale magnetic fields may form 
from small-scale magnetic fields through reconnection \citep{tou96}. The 
chaotic small-scale magnetic fields in the disk region are carried into the 
transition region by accreting gases, then form a large-scale magnetic field 
in the transition region through reconnection.
(3) The chaotic small-scale magnetic fields in the disk region are carried 
into the transition region by accreting gases, then stretched by the radial 
motion to increase the correlation length in the radial direction. Imagine
that, in the disk region and near the inner boundary, there is a small loop 
of magnetic field line with a length $\sim H$ in the radial direction, where 
$H$ is the half-thickness of the disk. As accretion goes on, its inner foot 
(the foot closer to the inner boundary) leaves the disk first, then moves 
to the black hole along with the gas. The outer foot is still in the disk 
region and has a distance $\sim H$ from the inner boundary of the disk, moves 
toward the inner boundary with a velocity $v_r$. Assuming that the time needed
by a plasma particle moving from the inner boundary of the disk to the horizon 
of the black hole is approximately given by the radial free-fall 
time:\footnote{In a real case a particle leaving from the inner boundary of
disk has nonzero angular momentum so will not free-fall radially. However, if 
the magnetic field is dynamically important in the transition region (as the 
case in the present paper), the particle may lose its angular momentum quickly 
so the precise proper time taken by the particle moving from the disk inner 
boundary to the black hole horizon may not be far from the radial free-fall 
time.} $t_1\sim
r_{\rm in} (r_{\rm in}/r_{\rm g})^{1/2}$. Then, the correlation length of the
magnetic field in the transition region is $\gtrsim r_{\rm in}$ if the following
condition is satisfied: $t_1 < t_2 \equiv H/v_r \sim \left(c_{\rm s}^2 / 
v_\phi^2\right) r_{\rm in} /v_r$, where $c_{\rm s}$ is the sound speed in the 
disk region, and $v_\phi$ is the rotation velocity of the disk. For a Keplerian 
disk we have $v_\phi \sim (r_{\rm g}/r)^{1/2}$, then $t_1 < t_2$ leads to $v_r 
< c_{\rm s}^2 / v_\phi$ in the disk. Interestingly, for a standard Keplerian 
disk we have $v_r \sim c_s^2 / v_\phi$ \citep{bal98}. So, if the disk radial 
velocity is smaller than that predicted by a standard Keplerian disk, then
a large-scale magnetic field can be formed in the transition region.

\section*{Acknowledgments}

The author thanks Ramesh Narayan and Bohdan Paczy\'{n}ski for advices and 
discussions, and the anonymous referee for helpful comments. This research 
was supported by NASA through Chandra Postdoctoral Fellowship grant number 
PF1-20018 awarded by the Chandra X-ray Center, which is operated by the 
Smithsonian Astrophysical Observatory for NASA under contract NAS8-39073.


\appendix

\section{Structures of the Solutions}
\label{app1}

In the main text we have solved the solutions at the fast critical points 
and the corresponding integral constants. However, this does not guarantee
the existence of global solutions that start from the inner boundary of
the disk and end at the horizon of the black hole. In this appendix we show
that global solutions exist corresponding to the fast critical point 
solutions in Fig.~\ref{fig1} except the branch B in the case of $a=0$, 
by presenting some examples of global solutions. (The inner boundary of the 
disk is always at the marginally stable circular orbit.)

\begin{figure}
\begin{center}
\FigureFile(12cm,){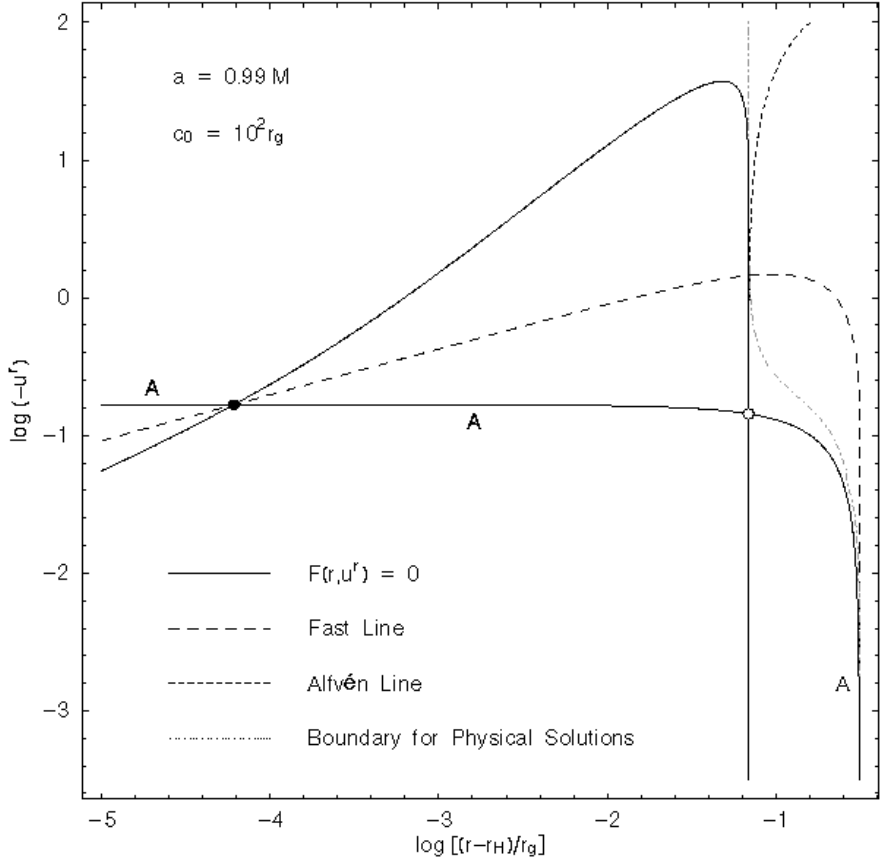}
\end{center}
\caption{The contour of $F(r,u^r) = 0$ (solid curves), corresponding to 
$a = 0.99 M$ and $c_0 = 10^2 r_{\rm g}$. The contours for the fast critical
points [the long-dashed curve, defined by $u_r u^r = c_{\rm A}^2/\left(1-  
c_{\rm A}^2\right)$], the Alfv\'en point (the short-dashed curve, defined by 
$u_r u^r = c_{{\rm A}r}c_{\rm A}^{~\,r}$), and the boundary for physical 
solutions (the dashed-dotted curve, beyond which $f_{\rm L}$ is complex so 
physical solutions do not exist) are also shown. The global
solution $u^r = u^r(r)$ is represented by the smooth solid curve labeled
with letter ``A'', which starts from the inner boundary of the disk (the right
end) and ends on the horizon of the black hole (the left end), passing the
Alfv\'en critical point (the circle) and the fast critical point (the filled
circle) in turn.
\label{fig7}}
\end{figure}

In Fig.~\ref{fig7}, we present the global solution $u^r = u^r(r)$ for $a =
0.99 M$ and $c_0 = 10^2 r_{\rm g}$. The corresponding fast critical point
is at $r_{\rm f} = 1.141 r_{\rm g}$ and $u_{\rm f}^r = -0.1673$, integral
constants are $f_{\rm L} = 1215 r_{\rm g}$ and $f_{\rm E} = 442.7$ (so the
total radiation efficiency is $\varepsilon_{\rm total} = 443.7$). In the 
figure, the solid curves correspond to $F(r,u^r) = 0$. The smooth solid 
curve going from the bottom-right corner to the top-left corner, as labeled 
with letter ``A'', represents the global solution for $u^r = u^r(r)$. The
solution describes a flow that starts from the marginally stable circular 
orbit subsonically (the right end), passes the Alfv\'en point (the circle) 
first, then the fast
critical point (the filled circle), finally falls into the horizon of the 
black hole supersonically (the left end). 

\begin{figure}
\begin{center}
\FigureFile(12cm,){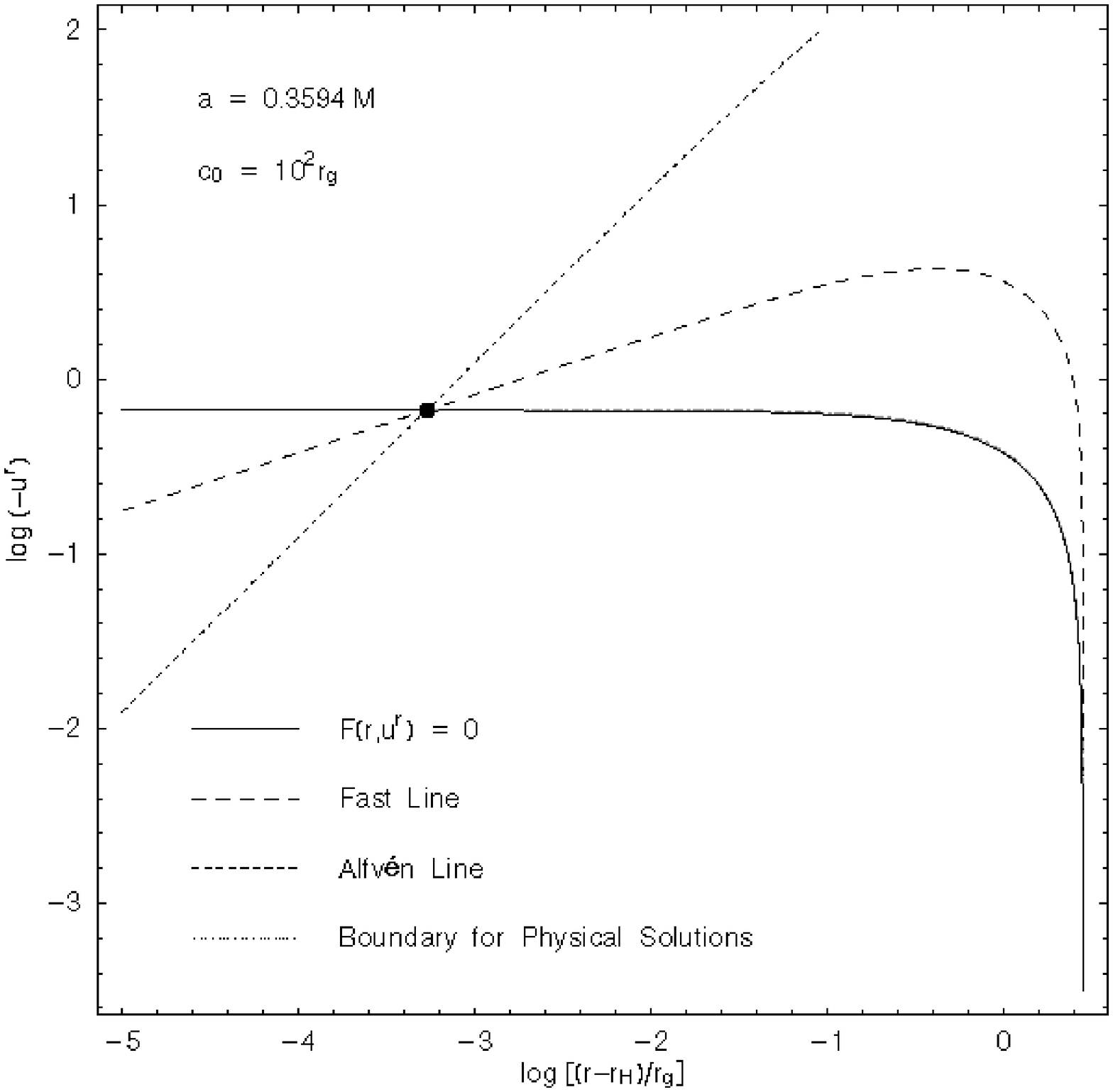}
\end{center}
\caption{Similar to Fig.~\ref{fig7} but for  $a = a_{\rm cr} =0.3594 M$. 
The Alfv\'en critical point merges to the fast critical point, and they are
right on the boundary for physical solutions (c.f. Fig.~\ref{fig1}, the
panel for $a =0.3594 M$). (The dashed-dotted curve is moved upward a little
bit so that it can be seen. Indeed it almost coincides with the solid curve.)
\label{fig8}}
\end{figure}

In Fig.~\ref{fig8} we show the global solution $u^r = u^r(r)$ for $a =
a_{\rm cr}$ and $c_0 = 10^2 r_{\rm g}$. The corresponding fast critical 
point is at $r_{\rm f} = 1.934 r_{\rm g}$ and $u_{\rm f}^r = -0.6627$, 
integral constants are $f_{\rm L} = -0.7556 r_{\rm g}$ and $f_{\rm E} = 
-0.7110$ (so the total radiation efficiency is $\varepsilon_{\rm total} = 
0.289$). Note, in this example, the fast critical point is almost right on 
the boundary for physical solutions (see Fig.~\ref{fig1}, the panel for 
$a = 0.3594 M$), and the Alfv\'en critical point merges to the fast critical 
point. Comparing Fig.~\ref{fig8} to Fig.~\ref{fig7}, we can see how the 
topology of the solutions to $F(r, u^r) =0$ changes as $a \rightarrow 
a_{\rm cr}$ from above.

\begin{figure}
\begin{center}
\FigureFile(12cm,){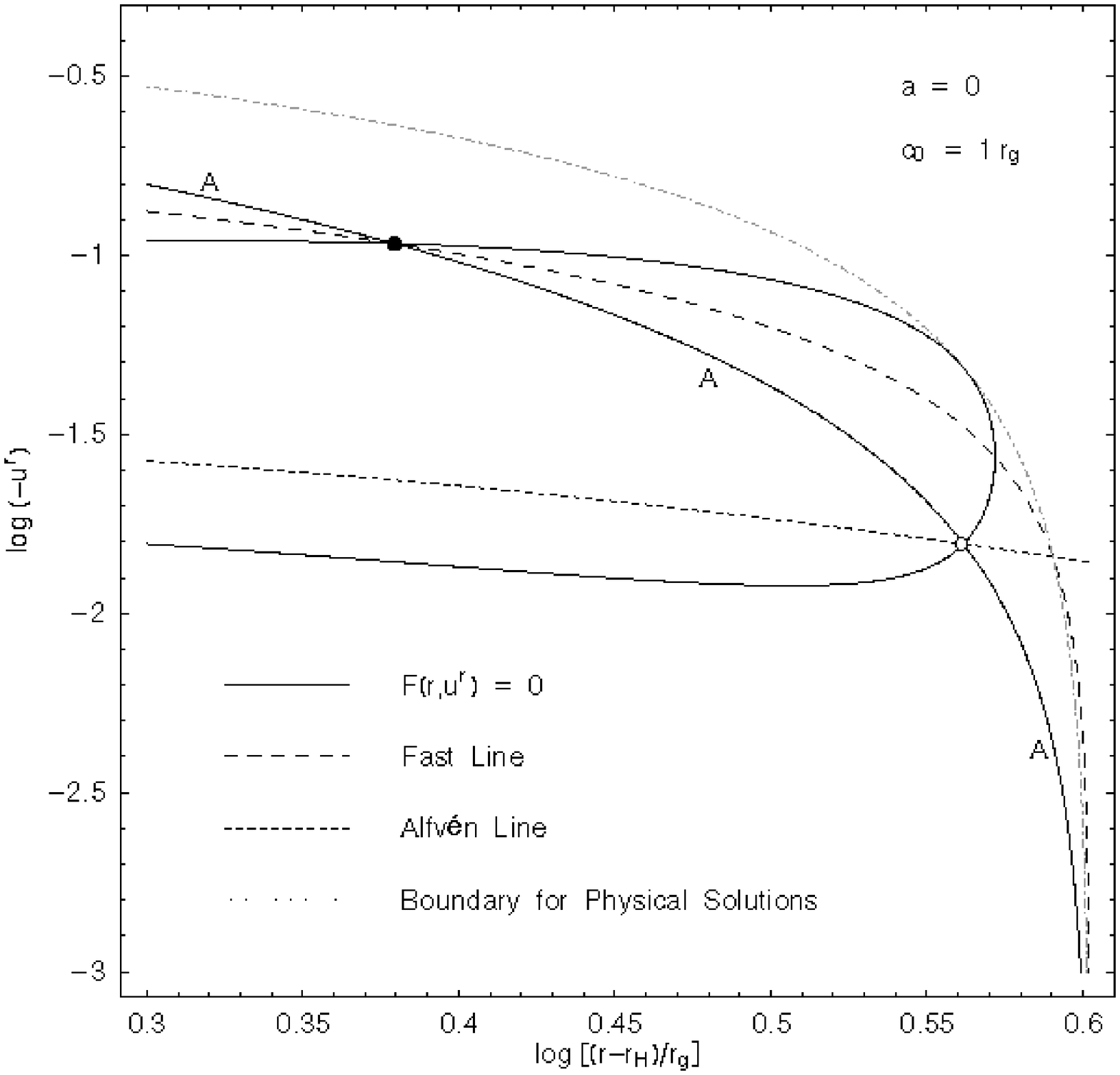}
\end{center}
\caption{Similar to Fig.~\ref{fig7} but for $a = 0$ and $c_0 = 1 r_{\rm g}$.
The smooth solid curve labeled with letter ``A'' is a physical solution for 
$u^r = u^r(r)$, starting from the inner boundary of the disk (the right end) 
and ending on the horizon of the black hole (the left end).
\label{fig9}}
\end{figure}

In Fig.~\ref{fig9} we show the global solution $u^r = u^r(r)$ for $a =
0$ and $c_0 = 1 r_{\rm g}$. The corresponding fast critical point is at 
$r_{\rm f} = 4.395 r_{\rm g}$ (on the branch A in Fig.~\ref{fig1}, the 
panel for $a=0$) and $u_{\rm f}^r = -0.1079$, integral constants are 
$f_{\rm L} = -3.078 r_{\rm g}$ and $f_{\rm E} = -0.9165$ (so the total 
radiation efficiency is $\varepsilon_{\rm total} = 0.083$). 

\begin{figure}
\begin{center}
\FigureFile(12cm,){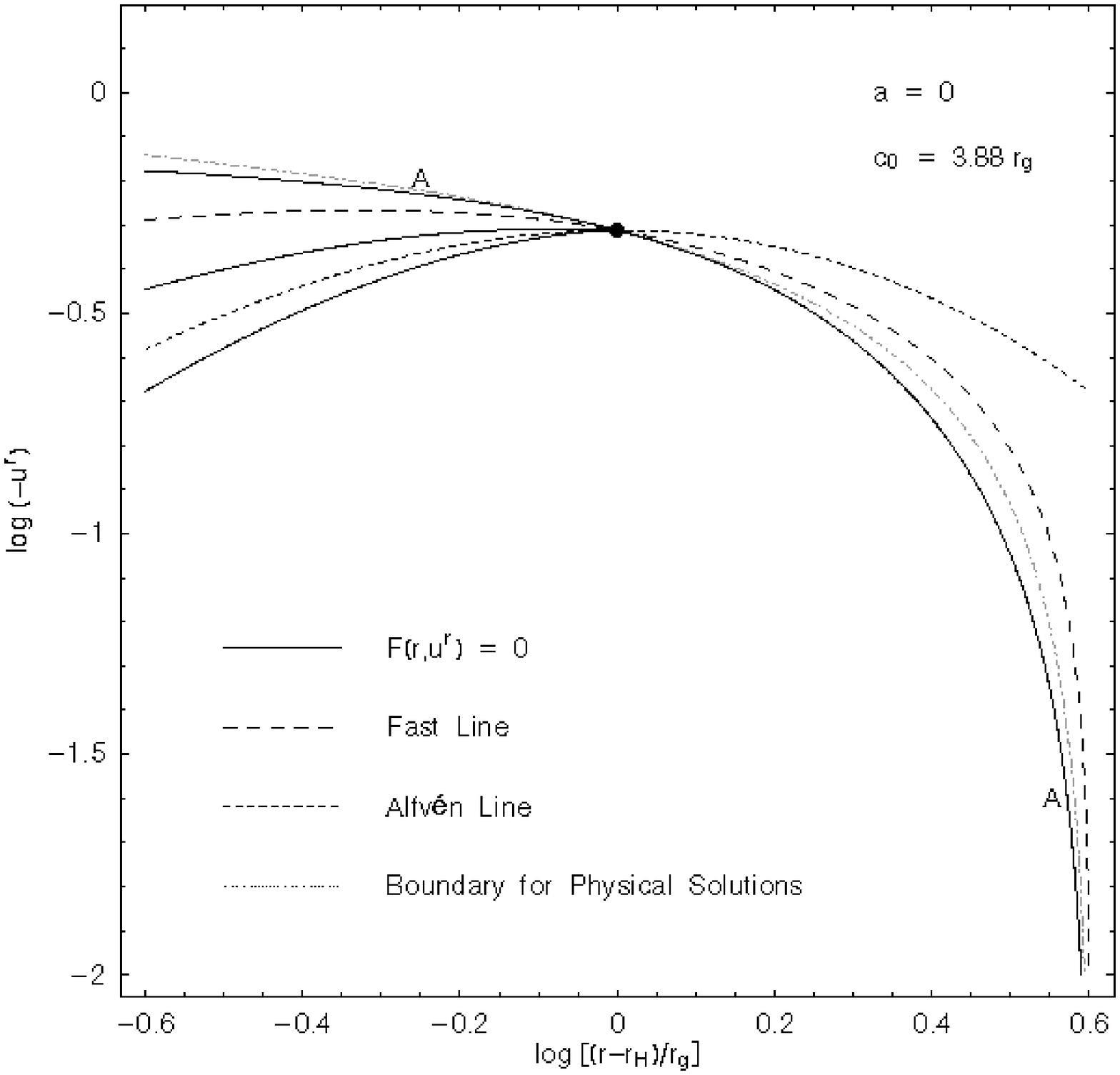}
\end{center}
\caption{Similar to Fig.~\ref{fig9} but for $c_0 = 3.88 r_{\rm g}$. In this
case the fast critical point is on the boundary for physical solutions (c.f. 
Fig.~\ref{fig1}, the panel for $a =0$), and the Alfv\'en critical point 
merges to the fast critical point. 
\label{fig10}}
\end{figure}

In Fig.~\ref{fig10} we show the global solution $u^r = u^r(r)$ for $a =
0$ and $c_0 = 3.88 r_{\rm g}$. The corresponding fast critical point is at 
$r_{\rm f} = 3.00 r_{\rm g}$ (the dark point on the branch A in 
Fig.~\ref{fig1}, the panel for $a=0$) and $u_{\rm f}^r = -0.488$, integral 
constants are $f_{\rm L} = -1.485 r_{\rm g}$ and $f_{\rm E} = -0.808$
(so the total radiation efficiency is $\varepsilon_{\rm total} = 0.19$).
Similar to Fig.~\ref{fig7}, the smooth solid curve labeled with letter  
``A'' is the physical solution that starts from the inner boundary of 
the disk and ends on the horizon of the black hole. Note, in this example, 
the fast critical point is on the boundary for physical solutions (c.f. 
Fig.~\ref{fig1}, the panel for $a = 0$), and the Alfv\'en critical point 
merges to the fast critical point. Comparing Fig.~\ref{fig10} to 
Fig.~\ref{fig9}, we can see how the topology of the solutions to $F(r, 
u^r) =0$ changes as the fast critical point approaches the boundary for 
physical solutions.

Figures~\ref{fig7}--\ref{fig10} demonstrate that global solutions exist
for $a\geq c_{\rm cr}$ with any $\alpha_c = c_0/r_{\rm g}$, and for $0\leq 
a<a_{\rm cr}$ with $\alpha_c \leq \alpha_{c,\max}$.

\begin{figure}
\begin{center}
\FigureFile(12cm,){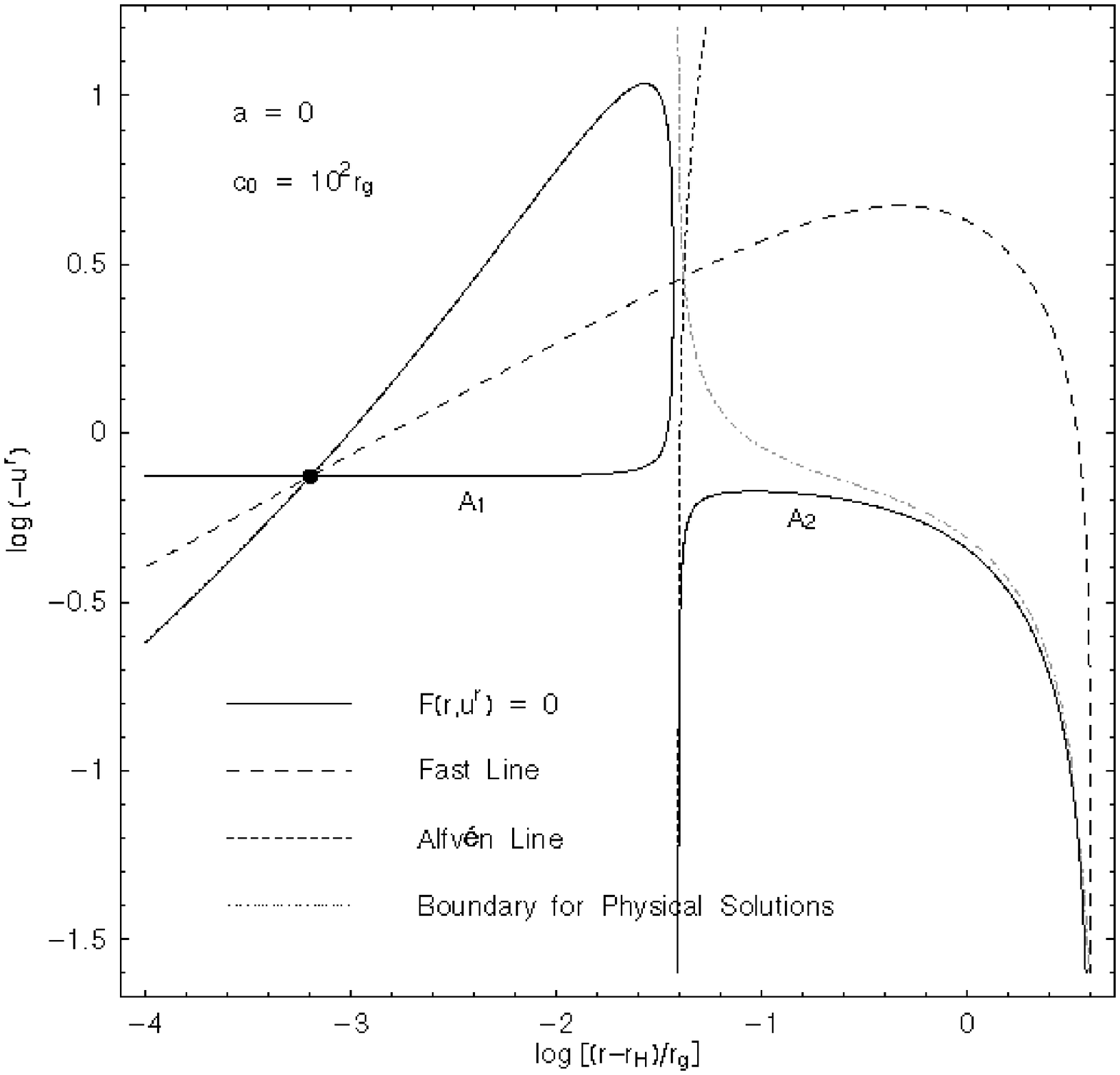}
\end{center}
\caption{The contour of $F(r,u^r) = 0$ (solid curves), corresponding to 
$a = 0$ and $c_0 = 10^2 r_{\rm g}$, and $r_{\rm f} = 2.00064 r_{\rm g}$. 
There is no physical solution corresponding to a flow that starts 
from the marginally stable circular orbit, passes the fast critical point,
then enters the horizon of the black hole. As shown, the curves A$_1$ and
A$_2$ are not connected to each other.
\label{fig11}}
\end{figure}

Finally, we show that global solutions do not exist for $0\leq a<a_{\rm 
cr}$ and $\alpha_c > \alpha_{c,\max}$, by presenting the contours of $F(r,
u^r) =0$ for $a = 0$ and $c_0 = 10^2 r_{\rm g}$ in Fig.~\ref{fig11}. The
corresponding solutions for the fast critical point is at $r_{\rm f} = 
2.00064 r_{\rm g}$ (on the branch B in Fig.~\ref{fig1}, the panel for 
$a=0$) and $u_{\rm f}^r = -0.74377$, integral constants are $f_{\rm L} = 
701.7 r_{\rm g}$ and $f_{\rm E} = 47.04$. As we can see from the figure,
there is no solid curve representing a solution $u^r = u^r(r)$ that starts
from the marginally stable circular orbit ($r_{\rm ms} = 6 r_{\rm g}$)
and ends on the horizon of the black hole ($r_{\rm H} = 2 r_{\rm g}$):
the curves A$_1$ and A$_2$ are not connected to each other. This conclusion 
can also be inferred from the values of $f_{\rm E}$: it is positive so 
cannot correspond to physical solutions since no energy can be extracted
from a Schwarzschild black hole. (On the branch B in Fig.~\ref{fig1},
the panel for $a = 0$, there is another solution for $r_{\rm f}$ 
corresponding to $c_0 = 10^2 r_{\rm g}$: $r_{\rm f} =2.0315 r_{\rm g}$.
It is easy to show that this also does not lead to a physical solution.)

\section{Some Useful Fitting Formulas}
\label{app2}

In this Appendix we present some useful fitting formulas. The inner boundary
of the disk is always assumed at the marginally stable circular orbit.

For $0.3594 <a/M <0.999$ and $c_0^2/r_{\rm g}^2\gg 1$, the solution for the
radius at the fast critical point can be fitted by
\begin{eqnarray}
     r_{\rm f} \approx r_{\rm H} + f_1(a^\star)\, r_{\rm g}^3 /c_0^2,
	\label{fit_rf}
\end{eqnarray}
where $a^\star\equiv a/M$ and 
\begin{eqnarray}
     f_1(a^\star) \equiv 3.637 + 2.319 \sqrt{1 -a^\star} + 1.672 a^\star
          -4.883 a^{\star 2}- 4.975 a^{\star 3}+ 4.882 a^{\star 4}.
	\label{fa}
\end{eqnarray}
The relative error in $r_{\rm f}-r_{\rm H}$ is $<6\%$ for $c_0/r_{\rm g} >10$, 
$<2\%$ for $c_0/r_{\rm g} >100$. Then, since $r_{\rm f} -r_{\rm H} \ll r_{\rm 
H}$, the radial component of the four-velocity at the fast critical point is
\begin{eqnarray}
     u_{\rm f}^r \approx - \left(\frac{M}{r_{\rm H}}\right)^{4/3}
	     \left\{2 f_1 (a^\star)\sqrt{1-a^{\star 2}} \left[f_{\rm E}^{\prime
		2}+ 4M^2 \left(\Omega_{\rm H}-\Omega_\Psi\right)^2\right]
		\right\}^{1/3}.\hspace{1.45cm}
	\label{fit_uf}
\end{eqnarray}

Substituting $r = r_{\rm f}$ and $u = u_{\rm f}^r$ into $F(r,u^r) =0$, we
can solve for $f_{\rm L}$, then $f_{\rm E} = f_{\rm E}^\prime +\Omega_\Psi
f_{\rm L}$.

For $0\leq a/M \leq 0.3952$, the maximum value of $c_0$ (beyond which 
stationary and axisymmetric solutions do not exist, i.e. the $c_0$ on the 
dashed curve in Fig.~\ref{fig5}), can be fitted by $\log (c_0/r_{\rm g})
\approx f_2(a^\star)$, where
\begin{eqnarray}
     f_2(a^\star) &\equiv& 0.5718 + 0.01069/ \sqrt{0.3594 - a^\star}
          -0.00006044/(0.3594 - a^\star) - 0.1928 a^\star \nonumber\\
		&& + 8.255 a^{\star 2}- 114.5 a^{\star 3} + 735.4 a^{\star 4} 
		- 2161 a^{\star 5} + 2402 a^{\star 6}.
\end{eqnarray}
The relative error in $\log (c_0/r_{\rm g})$ is $< 1\%$.

For $0.3956 \leq a/M \leq 0.999$, the $c_0$ corresponding to the ``equilibrium
state'' in Fig.~\ref{fig5} can be fitted by $\log (c_0/r_{\rm g}) \approx f_3
(a^\star)$, where
\begin{eqnarray}
     f_3(a^\star) &\equiv& 31.803 + 2.447\sqrt{1 - a^\star}
          + 0.01371/ \sqrt{a^\star-0.3594} - 297.945 a^\star 
          + 1104.82 a^{\star 2}\nonumber\\
          && - 2160.917 a^{\star 3} + 2357.443 a^{\star 4} 
          - 1359.334 a^{\star 5}+ 323.972 a^{\star 6}.
\end{eqnarray}
The relative error in $\log (c_0/r_{\rm g})$ is $< 2\%$.

\label{lastpage}

\end{document}